\documentclass[12pt]{article}
\usepackage{epsfig}
\usepackage{latexsym}
\usepackage[hyperindex]{hyperref}

\newskip\defaultbaselineskip\defaultbaselineskip=12pt

\def\GeV{\mathord{\rm \;GeV}}
\def\MeV{\mathord{\rm \;MeV}}

\def\bar{\overline}

\def\eqinf{\mathrel{\raise2pt\hbox to 24pt{\raise -8pt\hbox{$t \to \infty$}\hss{$=$}}}}

\def\gsim{\mathrel{\raise2pt\hbox to 8pt{\raise -5pt\hbox{$\sim$}\hss{$>$}}}}
\def\rsim{\mathrel{\raise2pt\hbox to 8pt{\raise -5pt\hbox{$\sim$}\hss{$>$}}}}
\def\lsim{\mathrel{\raise2pt\hbox to 8pt{\raise -5pt\hbox{$\sim$}\hss{$<$}}}}
\def\ssqr#1#2{{\vbox{\hrule height.#2pt
      \hbox{\vrule width.#2pt height#1pt \kern#1pt\vrule width.#2pt}
      \hrule height.#2pt}\kern-.#2pt}}

\def\Dsl{\,\raise.15ex\hbox{$/$}\mkern-13.5mu D} %this one can be subscripted
\def\dsl{\raise.15ex\hbox{$/$}\kern-.57em\hbox{$\partial$}}

\def\CA{{\cal A}}  \def\CB{{\cal B}}  \def\CC{{\cal C}} 
  \def\CF{{\cal F}}  \def\CF{{\cal F}} 
     
\def\CM{{\cal M}}    \def\CO{{\cal O}} \def\CP{{\cal P}}
\def\CQ{{\cal Q}}    \def\CS{{\cal S}} \def\CT{{\cal T}}
  \def\CV{{\cal V}}

\ifx\href\undefined\def\href#1#2{{#2}}\fi
\def\spireshome{http://www.slac.stanford.edu/cgi-bin/spiface/find/hep/www?FORMAT=WWW&}
{\catcode`\%=12
\xdef\spiresjournal#1#2#3{\noexpand\protect\noexpand\href{\spireshome
                          rawcmd=find+journal+#1%2C+#2%2C+#3}}
\xdef\spireseprint#1#2{\noexpand\protect\noexpand\href{\spireshome rawcmd=find+eprint+#1%2F#2}}
\xdef\spiresreport#1{\noexpand\protect\noexpand\href{\spireshome rawcmd=find+rept+#1}}
\xdef\spireskey#1{\noexpand\protect\noexpand\href{\spireshome key=#1}}
}
\def\eprint#1#2{\spireseprint{#1}{#2}{#1/#2}}
\def\report#1{\spiresreport{#1}{#1}}
\def\nohref{}

\def\putpaper{\edef\refpage{\the\count0}%
              \def\nohref{}%
              {\def\ {+}\def\nohref##1{}\edef\temp{\noexpand\spiresjournal
               {\journalname}{\volume}{\refpage}}\expandafter}\temp
               {\sfcode`\.=1000{\journalname} {\bf \volume} (\refyear)
                \refpage}\egroup}
\def\putpage{\edef\refpage{\the\count0}%
              \def\nohref{}%
              {\def\ {+}\def\nohref##1{}\edef\temp{\noexpand\spiresjournal
               {\journalname}{\volume}{\refpage}}\expandafter}\temp
              {\refpage}\egroup}
\def\dojournal#1#2 (#3){\def\journalname{#1}\def\volume{#2}\def\refyear
                        {#3}\afterassignment\putpaper\bgroup\count0=}
\def\morepage{\afterassignment\putpage\bgroup\count0=}

\def\NPB#1{\dojournal{Nucl.\ Phys.}{B#1}}
\def\NPBPS#1{\dojournal{Nucl.\ Phys.\ \nohref(Proc.\ Suppl.\nohref)}{\nohref B#1}}

\def\PRL#1{\dojournal{Phys.\ Rev.\ Lett.}{#1}}

\def\PRD#1{\dojournal{Phys.\ Rev.}{D#1}}

\def\PLB#1{\dojournal{Phys.\ Lett.}{#1B}}

\def\ZPC#1{\dojournal{Z.\ Phys.}{C#1}}

\def\etal{{\it et al.}}

\def\gsim{\mathrel{\raise2pt\hbox to 8pt{\raise -5pt\hbox{$\sim$}\hss{$>$}}}}
\def\rsim{\mathrel{\raise2pt\hbox to 8pt{\raise -5pt\hbox{$\sim$}\hss{$>$}}}}
\def\lsim{\mathrel{\raise2pt\hbox to 8pt{\raise -5pt\hbox{$\sim$}\hss{$<$}}}}
\def\half{{\textstyle{1\over2}}} \def\third{{\textstyle{1\over3}}}

\def\vev#1{\langle #1\rangle}
\def\nfrac#1#2{{\textstyle{#1\over#2}}} 

\newcommand\figcaption[1]{\vskip-0.0truein\caption{#1}\vskip0.0truein}

\newcommand\NDR{{\hbox{\rm NDR}}}
\newcommand\MSbar{{\hbox{$\overline{MS}$}}}
\newcommand\mbar{\hbox{$\overline{m}$}}

\newcommand{\newsection}[1]{\section{#1}\setcounter{equation}{0}}

\setlength\hfuzz{3pt}
\hbadness=5000

\begin{document}

%% \title{Matrix elements of 4-fermion operators and B-parameters 
%%        with Wilson Fermions}
%% 	
%% \begin{center}
%% R.~Gupta and Tanmoy Bhattacharya
%% \vskip 0.3 truein
%% Group T-8, MS B285, Los Alamos National Laboratory, Los Alamos, New Mexico 87545 U.~S.~A.
%% \vskip 0.5 truein
%% S.~Sharpe
%% \vskip 0.2 truein
%% Physics Department, University of Washington, Seattle, WA 43210.
%% \end{center}
%% 
%% \maketitle

\begin{titlepage}
 \null
 \begin{center}
 \makebox[\textwidth][r]{LAUR-96-1829}
 \makebox[\textwidth][r]{UW/PT-96-12}
 \par\vspace{0.25in} %%%
  {\Large
      MATRIX ELEMENTS OF 4-FERMION OPERATORS \\ [0.5em]
      WITH QUENCHED WILSON FERMIONS}
  \par
 \vskip 2.0em
 
 {\large 
  \begin{tabular}[t]{c}
        Rajan Gupta\footnotemark 
	and Tanmoy Bhattacharya\footnotemark\\[0.5em]
        \em Group T-8, Mail Stop B-285, Los Alamos National Laboratory\\
        \em Los Alamos, NM 87545, U.~S.~A\\[1.5em]
        Stephen R. Sharpe \footnotemark\\[0.5em]
        \em Physics Department, Box 351560, University of Washington \\
        \em Seattle, WA 98195-1560 \\
  \end{tabular}}
 \par \vskip 2.0em
 {\large\bf Abstract}
\end{center}

\quotation We presents results for the matrix elements of a variety of
four-fermion operators calculated using quenched Wilson fermions.  Our
simulations are done on 170 lattices of size $32^3 \times 64$ at
$\beta = 6.0$. We find $B_K=0.74 \pm 0.04 \pm 0.05$, $B_D= 0.78 \pm
0.01$, $ B_7^{3/2}= 0.58 \pm 0.02 {+0.07 \atop -0.03}$, $ B_8^{3/2}=
0.81 \pm 0.03 {+.03 \atop -0.02}$, with all results being in the NDR
scheme at $\mu=2\GeV$.  We also calculate the B-parameter for
%%% NEW
the operator $\CQ_s$, which is needed in the study of the difference
of B-meson lifetimes.  Our best estimate is $B_S(\NDR,\mu=1/a=2.33
\GeV)=0.80 \pm 0.01$. This is given at the lattice scale since the
required 2-loop anomalous dimension matrix is not known.  In all these
estimates, the first error is statistical, while the second is due to
the use of truncated perturbation theory to match continuum and
lattice operators.  Errors due to quenching and lattice discretization
are not included.  We also present new results for the perturbative
matching coefficients, extending the calculation to all Lorentz scalar
four-fermion operators, and using NDR as the continuum scheme.

\footnotetext[1]{Email: rajan@qcd.lanl.gov}
\footnotetext{Email:    tanmoy@qcd.lanl.gov}
\footnotetext{Email:    sharpe@phys.washington.edu}
\vfill
\mbox{NOV 21, 1996}
\end{titlepage}

\setlength{\textfloatsep}{12pt plus 2pt minus 2pt}

\makeatletter % Thus allowing override of non-user parameters

% We do not want extra spacing between items
\setlength{\leftmargini}{\parindent}
\def\@listi{\leftmargin\leftmargini
            \topsep 0\p@ plus2\p@ minus2\p@\parsep 0\p@ plus\p@ minus\p@
            \itemsep \parsep}
% Do not want line break after table number in table captions.
\long\def\@maketablecaption#1#2{#1. #2\par}

% Allow a tiny stretch between paragraphs
\advance \parskip by 0pt plus 1pt minus 0pt

\makeatother

\newsection{Introduction}
\label{sec:intro}

One of the central goals of lattice QCD is to calculate hadronic
matrix elements of phenomenological interest 
\cite{lusignoli,burasREV}.
We present here results for the matrix elements of 
a variety of four-fermion operators:
$B_K$, which is needed as input for estimates of CP violation in 
$K-\bar{K}$ mixing;
$B_D$, required to estimate $D-\bar{D}$ mixing, 
and an indication of the result for $B$ mesons;
$B_7^{I=3/2}$ and $B_8^{I=3/2}$, which determine the dominant
contribution of electromagnetic penguins to $\epsilon'/\epsilon$;
and a new quantity, $B_S$, needed as part of an estimate of
the lifetime difference between $B$ mesons \cite{beneke}.

Our results are obtained using Wilson fermions in
the quenched approximation.
The use of Wilson fermions makes it more difficult
to extract quantities which are constrained by chiral symmetry.
This is a problem for the calculation of $B_K$, and our 
results are not competitive with those from staggered fermions.
Our aim here is to gain experience at reducing systematic errors.
The other matrix elements are not constrained by chiral symmetry,
and for these it is more straightforward to use Wilson
than staggered fermions.

Compared to previous work with Wilson fermions, our study uses
larger lattices ($32^3\times 64$), which should make finite volume effects
negligible and reduces the statistical errors,
has improved statistics (170 lattices), 
and uses a larger range of light quark masses 
($0.3 m_{s,\rm phys}$--$2 m_{s,\rm phys}$).
The latter improvement allows us to do more reliable chiral extrapolations,
since we can include terms of $O(m_K^4)$.
We have results at only a single lattice spacing, $\beta\equiv 6/g^2=6$,
so we cannot extrapolate to the continuum limit of the quenched theory.
What we provide is one point with statistical errors small enough that
systematic errors due to chiral extrapolations and the truncation of
perturbation theory can be quantified.

As an adjunct to our numerical results, we have extended previous
calculations of the perturbative matching coefficients between lattice
and continuum operators.  We also present a renormalization
group improved matching formula which incorporates tadpole improved
lattice perturbation theory.

The plan of the paper is as follows.
We begin, in Sec.~\ref{sec:summary}, by summarizing our results.
In Sec.~\ref{sec:details} we describe the various technical details pertinent 
to this calculation.  Section~\ref{sec:matching} 
summarizes the matching between lattice and continuum operators.
Results for the various $B$-parameters are given in 
%%% NEW
the final four sections.
%\ref{sec:BKresults}-\ref{sec:BSresults}
We discuss two technical issues in appendices.
Appendix A describes a renormalization group improved matching formula,
while in Appendix B we present general formulae
for the one-loop matching coefficients for all Lorentz scalar
four-fermion operators between the lattice and the NDR scheme.

\newsection{Summary of results}
\label{sec:summary}

We begin with our result for $B_K$ (defined in Eq.~\ref{eq:bkdef}):
\begin{equation}
B_K(NDR,2\GeV) = 0.74 \pm 0.04 ({\rm stat}) \pm 0.05 ({\rm subt}) \,.
\end{equation}
The second error is an estimate of the uncertainty due to our
method of subtracting chiral artifacts.
This is a considerable improvement over our previous 
work \cite{Bk93LANL}---the 
statistical error has been reduced by a factor of $\sim 6$, 
%% comparing the U_1 numbers: 0.57(23) with the above
%% the reduction in errors in the bare numbers is 15 at p=0!
and we have better control over the systematic errors. 

It is interesting to compare our result to those other recent high
statistics calculations at or near the same lattice spacing,
particularly since all use different subtraction methods.  
Bernard and Soni use Wilson fermions at
$\beta=6$ on $24^3\times39$ lattices \cite{sonilat95}, and find
$B_K=0.67 \pm 0.07$.  Their subtraction method uses off-shell matrix
elements~\cite{Bk89SONI} whereas ours uses on-shell matrix elements at
finite momentum transfer.  One concern with their result is that they
use larger quark masses, which, according to our analysis, may lead to
an overestimate of $B_K$.  Nevertheless, it is encouraging that the
results agree.

The APE group has used improved ``clover'' fermions and a
non-perturbative determination of the matching coefficients 
\cite{Bk95APE,talevi}.
Their non-perturbatively matched operator has the correct chiral behavior.
They present their result 
(from $\beta=6$ on $18^3\times64$ lattices)
in the ``regularization independent'' scheme,
$B_K(RI,2.02 \GeV) = 0.62(11)$. 
The numerical value is unchanged upon conversion to NDR,
and is consistent with our result.
The two results can differ by terms of $O(a)$,
but the errors are too large to resolve any difference.

The JLQCD collaboration has recently presented results using Wilson
fermions and a non-perturbative determination of matching coefficients
\cite{kuramashi}. They find, at $\beta=5.9$ and $6.1$, that
$B_K(NDR,2\GeV)=0.48 \pm 0.05$ and $0.70 \pm 0.07$, respectively.
They also have a result for the value in the continuum limit,
$B_K(NDR,2\GeV,a=0)=0.59 \pm 0.08$.
The latter two numbers are consistent with our result, while the former is not. 
Our methods seem to work comparably well---the final errors are similar
based on similar samples and lattice sizes.
%($300$ of size $24^3\times64$ at $\beta=5.9$
%and $100$ of size $32^3\times64$ at $6.1$)
It will be interesting to use their non-perturbative
matching coefficients with our data to evaluate the accuracy of the 
procedure we use to remove the lattice artifacts in the chiral expansion. 

%The JLQCD collaboration has also extrapolated the lattice data 
%with perturbatively determined renormalization constants to the continuum limit. 
%Again they find $B_K(NDR,2\GeV,a=0)=0.59 \pm 0.08$, 
%in agreement with their
%extrapolated staggered value $B_K(NDR,2\GeV,a=0)=0.587 \pm 0.007 \pm 0.017$
% \cite{kuramashi}. 
%These comparisons show that while Wilson fermions have larger artifacts 
%at finite lattice spacing, they do give consistent results after 
%extrapolation to $a \to 0$.         

Finally, we compare our result with that from staggered
fermions. We expect that the two should agree up to corrections of $O(a)$.
At $\beta=6.0$, different choices of discretized staggered operators
give results in the range $0.68-0.71$
\cite{Bkstag90,sharpebk94,jlqcdbk,jlqcd96}, 
with statistical errors of $0.01$ or smaller. 
Our results are consistent with these, but our errors are too large
to allow us to see the expected $O(a)$ differences.

%\subsection{Summary of $B_D$ results}
\bigskip

Our final results for $B$-parameters for $D$ mesons
(defined in Eq.~\ref{eq:bddef}) are
\begin{eqnarray}
B_D(\NDR,\mu=2\GeV) &= & 0.785 \pm 0.015 \,,\\
B_{D_s}(\NDR,\mu=2\GeV) &=& 0.83 \pm 0.01 \,,\\
B_{D_s}/B_D &=& 1.047 \pm 0.014 \,.
\end{eqnarray}
The errors are much smaller than for $B_K$ as we need make no corrections
for mixing with wrong chirality operators. Thus, for this calculation,
we are at a point where the remaining systematic errors (finite
lattice spacing and quenching) are likely to be larger than the
statistical errors.  
%The variation of $B_D$ with the light quark mass
%is described by $0.777(15) + 0.77(27) m_q a$. Here $m_q a$ is the
%quark mass obtained from the Ward identity for the axial current, and
%$m_s$ is fixed using $M_\phi$ \cite{HM95LANL}.  
These results are consistent with those previously obtained
(see, for example, the compilation in Refs.~\cite{sonilat95} and \cite{flynn96}),
but have considerably smaller errors.

%Bernard and Soni find that this ratio decreases as the mass of the
%heavy quark is increased. Their latest result for $B-$mesons,
%$B_{bs}(2 \GeV) = B_{bd}(2 \GeV) = 1.02(13)$, is consistent with unity
%\cite{sonilat96}.  We do not have independent information on this
%variation with the mass of the heavy quark.  A recent review of the
%lattice results, extrapolated to the bottom quark mass, by the
%different groups can be found in \cite{flynn96}.

\bigskip

Our final results for the $B$-parameters of the $I=3/2$ parts of the
operators $\CQ_7$ and $\CQ_8$ (defined in Eqs.~\ref{eq:Q7}-\ref{eq:b8def})
 are
\begin{eqnarray}
  B_7^{3/2}(\NDR,2\GeV) &=&  0.58 \pm 0.02 ({\rm stat})
				{+0.07 \atop -0.03} ({\rm pert}) \,,  \\
  B_8^{3/2}(\NDR,2\GeV) &=&  0.81 \pm 0.03 ({\rm stat}) 
				{+0.03 \atop -0.02} ({\rm pert}) \,.
\end{eqnarray}
The ``perturbative error'' reflects the dependence of the results on the choice
of $\alpha_s$ used in the matching of continuum and lattice operators,
and is comparable to or larger than the statistical errors.
The perturbative error could be removed by the use of
non-perturbative matching coefficients, 
but these are not yet available for $\CQ_7$ and $\CQ_8$.
Our results are consistent with those we found
previously~\cite{Bk93LANL}, the apparent difference being due to the
use of a different continuum regularization scheme, a different final scale,
and a different choice of $\alpha_s$ in the matching of continuum and
lattice operators.

Our values are smaller than the numbers used by Lusignoli \etal
\cite{lusignoli} and Ciuchini \etal \cite{guidoBK} in their analyses
of predictions for $\epsilon'/\epsilon$: 
they take $B^{3/2}_7 = B^{3/2}_8 = 1.0 \pm 0.1$.  
This difference is important because a smaller $B_8$ means a
larger $\epsilon'/\epsilon$.

%\subsection{Summary for $B_S$}
\bigskip

The final quantities we consider are $B_S \equiv B_4^+$
and the related parameter $B_5^+$ 
[defined in Eqs.~\ref{eq:bsdef} and \ref{eq:b5def}].
What is of interest for phenomenology is the value of
$B_S$ for $\bar{b} s$ mesons.
The closest we can come is the result for b quark with roughly the 
mass of the charm, which is 
\begin{eqnarray}
  B_4^{+}(\NDR,1/a) &=&  0.80 \pm 0.01 ({\rm stat}) \,,  \\
  B_5^{+}(\NDR,1/a) &=&  0.94 \pm 0.01 ({\rm stat}) \,.
\end{eqnarray}
We have to quote these results at $\mu=1/a=2.33 \GeV$ because the two-loop
anomalous dimension matrix needed to run to $2\;$GeV has not been calculated.
For the same reason, we cannot estimate a ``perturbative'' error.

There are no previous results for these $B$-parameters using
propagating heavy quarks. Recently, Gim\'emez and Martinelli have
calculated $B_S$ in the static limit, $m_b\to\infty$~\cite{gimenez}.
Their lattice result, also at $\beta=6.0$ but without including any
perturbative corrections, 
is $B_S^{GM} = -(5/8) B_S = - (0.60 \pm 0.01 \pm 0.03)$.  
%%% NEW
While this is $\approx 20\%$ larger than our result,
the above mentioned differences preclude any meaningful comparison.

\newsection{Technical Details}
\label{sec:details}

We use a sample of 170 gauge configurations of size $32^3\times64$
generated at $\beta=6$ in the quenched approximation.
In the following we outline the method we have used to 
calculate matrix elements of four-fermion operators.
Further details of our update and inversion algorithms,
and of our determination of quark masses, can be found 
in Ref. \cite{HM95LANL}.

\subsection{Quark propagators}

In our approach we need two kinds of quark propagators:
one which allows for the
creation of mesons with an explicit zero-momentum projection, and the
other that allows good overlap with a range of lattice momenta.  
For the former we use wall sources (on a time slice fixed to Coulomb gauge);
for the latter, gauge invariant Wuppertal sources.
Both are calculated using periodic boundary conditions in all directions.
We calculate propagators at five quark masses:
$\kappa = 0.135$ ($C$), $0.153$ ($S$), $0.155$ ($U_1$), $0.1558$
($U_2$), and $0.1563$ ($U_3$).
These quark masses correspond to
pseudoscalar mesons of mass $2835$, $983$, $690$, $545$ and $431$
$\MeV$ respectively. Here, as in the following, we have
used $1/a=2.330(41)\GeV$, the scale we determined in Ref.~\cite{HM95LANL}
using $M_\rho$.  

With these five flavors we construct 15 distinct ``kaons'', whose
matrix elements we then study. The three $U_i$ quarks allow us to
extrapolate to the physical isospin symmetric light quark mass
$\mbar=(m_u+m_d)/2$, whose value is determined by matching
$M_\pi/M_\rho$ to experiment.  The physical value of strange quark
mass, determined using $M_\phi/M_\rho$, lies between $S$ and $U_1$,
and we use these two points to interpolate to it.  It turns out that
other ways of determining $m_s$, e.g. matching to $M_\pi^2/M_K^2$,
lead to values differing by $\sim 20\%$.  To estimate the uncertainty
that this introduces, we calculate kaon matrix elements in two ways:
by interpolating to our standard $m_s$, and by directly using the
results for $U_2 U_3$, which turns out to have almost exactly the
physical kaon mass, albeit for almost degenerate quarks. For $B_K,
B_7^{3/2}$, and $B_8^{3/2}$, the difference between the two results turn
out to be much smaller than other errors, so we do not quote a
systematic error due to the uncertainty in setting $m_s$.  In case of
$B_{D_s}$ and $B_S$, the difference between using $m_s(M_K)$ and
$m_s(M_\phi)$ is $\approx 1\%$ with $m_s(M_\phi)$ giving the larger
value.  For reasons explained in \cite{HM95LANL}, we believe that
$m_s(M_\phi)$ gives a better estimate of $m_s$, and we use this value
consistently for our final results.  To illustrate the details of our
analysis we often use results from $U_2 U_3$ as they are very close to
the physical kaon mass.

Finally, we take the physical charm mass to be $C$.
With this choice
the experimental values of $M_D,\ M_{D^*}$ and $M_{D_s}$ lie
between the static mass $M_1$ (measured from the rate of exponential
fall-off of the 2-point function) and the kinetic mass defined as $M_2
\equiv (\partial^2 E / \partial p^2 |_{p=0})^{-1}$ \cite{HM95LANL}.  

\subsection{Extracting the four-fermion matrix elements}

\begin{figure} %1
\hbox{\epsfxsize=\hsize\epsfbox{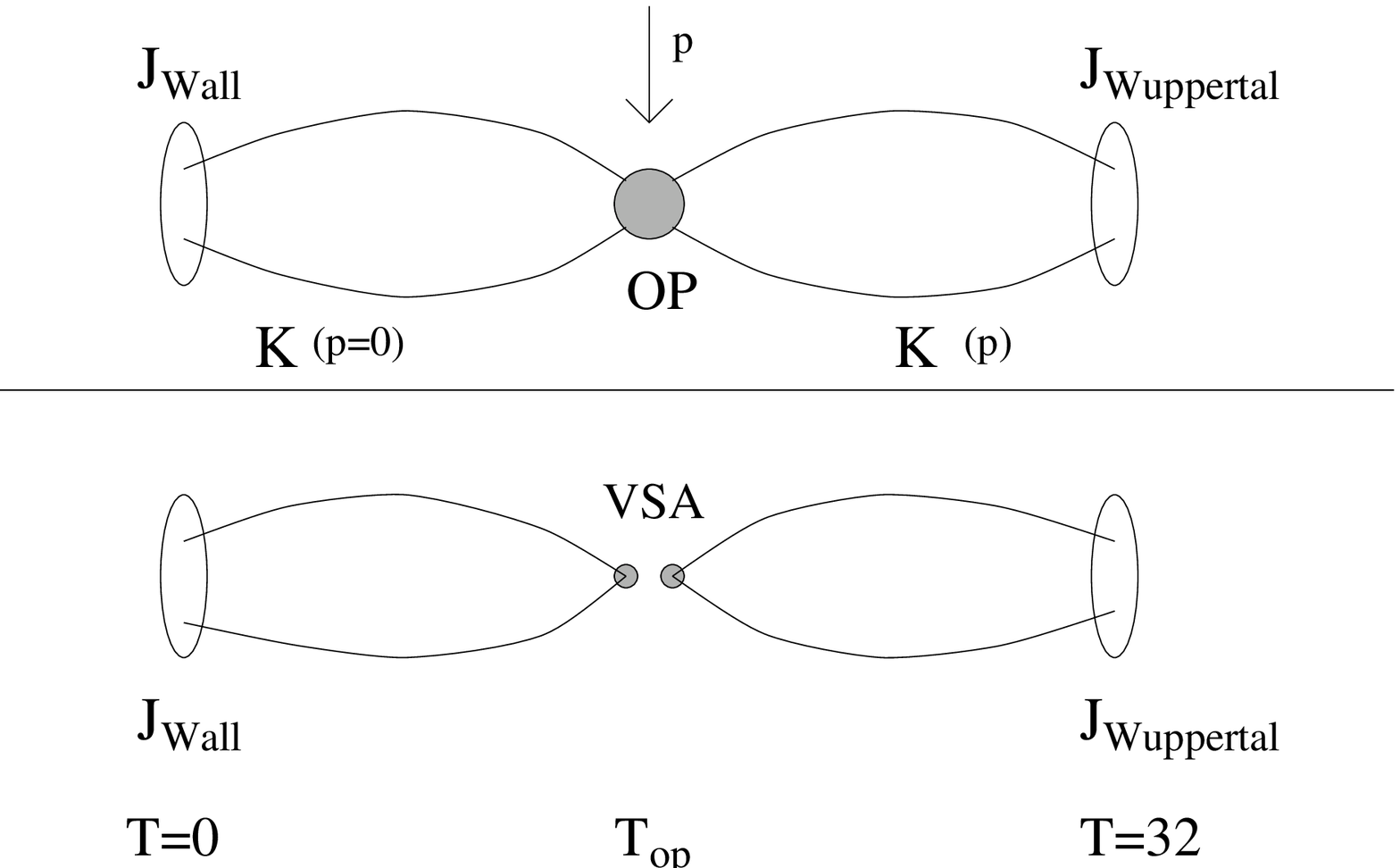}}
\figcaption{The ratio of correlations functions that we calculate.
The sources $J$ are the same for all four pseudoscalar mesons.
We show the picture for the case of propagation forward in time.}
%\vskip -24pt plus 10pt
\label{f_diagram}
\end{figure}

We use the same method for calculating matrix elements of four-fermion
operators as in our earlier work \cite{Bk93LANL}. 
The method is illustrated in Fig.~\ref{f_diagram}.
The initial pseudoscalar state is created by wall sources 
at time $T=0$, and thus has $\vec p = (0,0,0)$.
It propagates both forward and backwards in time.  
The four-fermion operator is inserted at a time in the range
$1-31$ ($63-33$ for the backward moving particle).
The operator
insertion is done at the 5 lowest lattice momenta, ${\vec p}=(0,0,0)$,
$(1,0,0)$, $(1,1,0)$, $(1,1,1)$ and $(2,0,0)$,
averaging over all possible permutations of the components of $\vec p$.  
The pseudoscalar emerging from the four-fermion operator is then
destroyed by an operator constructed using Wuppertal sources at
$T=32$. It is essential that this operator have a large coupling
to kaons with all the above momenta, and we find that Wuppertal
smearing does this well. 
In the end, we have two measurements,
corresponding to the forward and backward propagation, and, furthermore,
each of these is an average over a certain number of time slices.

As shown in Fig~\ref{f_diagram}, we calculate the ratio of the matrix
element of the four fermion operator to a product of bilinear matrix
elements.  The bilinears are either the time component of the axial
current or the pseudoscalar density.  Each of these bilinear matrix
elements is separately averaged over the gauge configurations.  In the
following we generically refer to the ratios of matrix elements as
$B$-parameters.  Using ratios cancels both the exponential decay
factors and the overlap of the source operators
%$J_{\rm Wall}$ and $J_{\rm Wuppertal}$ 
with the kaons.  This leads to simplified fitting, and to a reduction
in statistical errors.  For the renormalized $\Delta S=2$ LL operator,
the $B$-parameter one obtains (using the axial current in the
denominator) is proportional to $B_K$.  To calculate $B_7$ and
$B_8$ and $B_S$ is more complicated, since, as explained below, the
denominator involves both axial and pseudoscalar bilinears.

In several applications we need to consider the matrix element of the
four fermion operator at intermediate stages of the calculations. This
we obtain by multiplying the ratio by the product of bilinear matrix
elements, themselves calculated from 2-point functions.

%% Created using xfig on 4/7/96

\begin{figure} %2
\hbox{\epsfxsize=\hsize\epsfbox{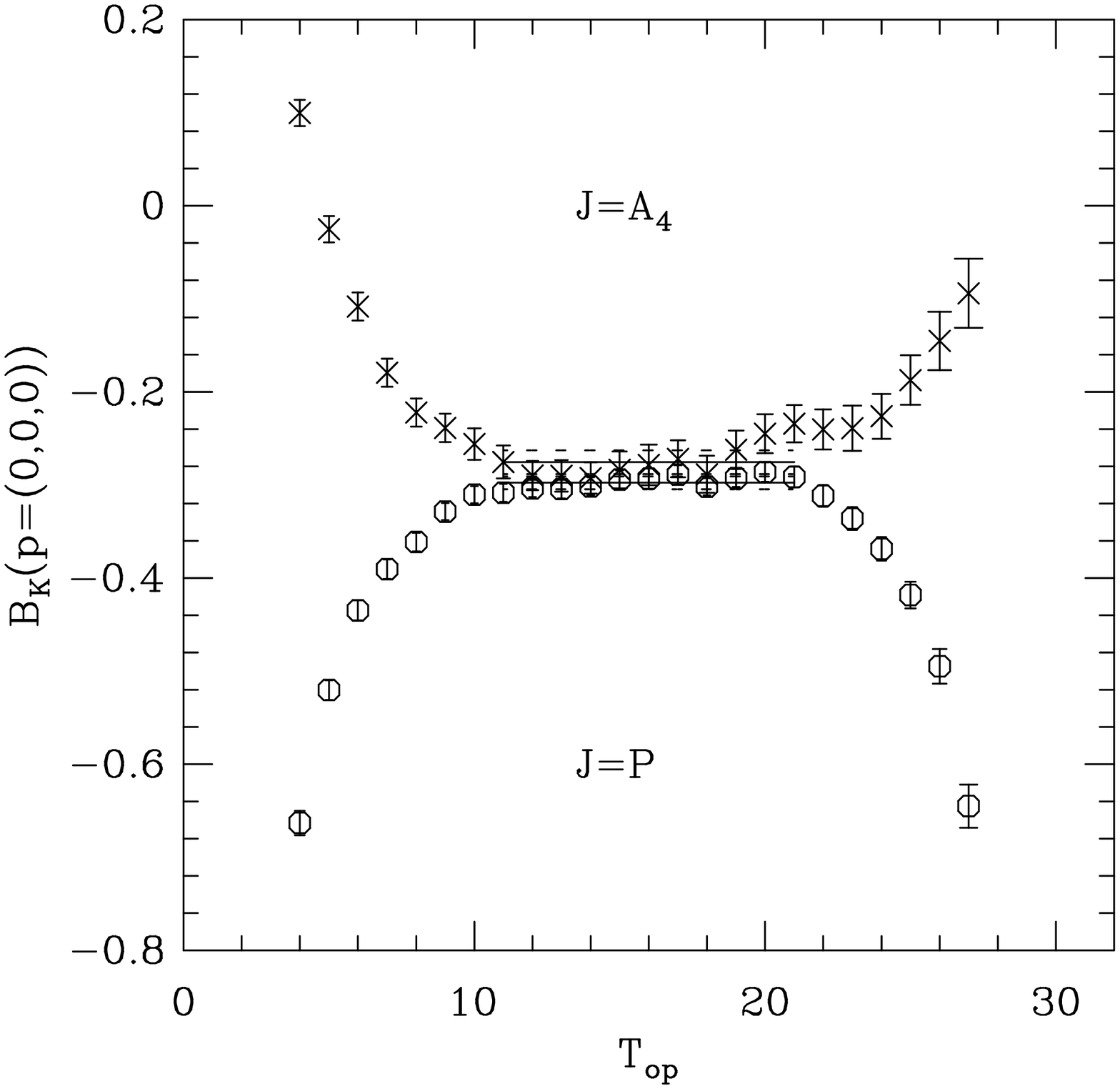}}
\figcaption{$B_K$ (using one-loop matching with $\mu=q^*=1/a$)
as a function of the time slice $T_{op}$ at which 
the operator is inserted for the two choices of source $J$. 
The data are for the mass combination $U_2U_3$ with momentum 
transfer $\vec p = (0,0,0)$. The fits are
shown by the solid line and the errors by the dashed lines.
The fit range, $T_{op}=11 - 21$, is chosen to be 
midway between the two sources.}
%\vskip -24pt plus 10pt
\label{f_bk_p0_u2u3}
\end{figure}

\begin{figure} %3
\hbox{\epsfxsize=\hsize\epsfbox{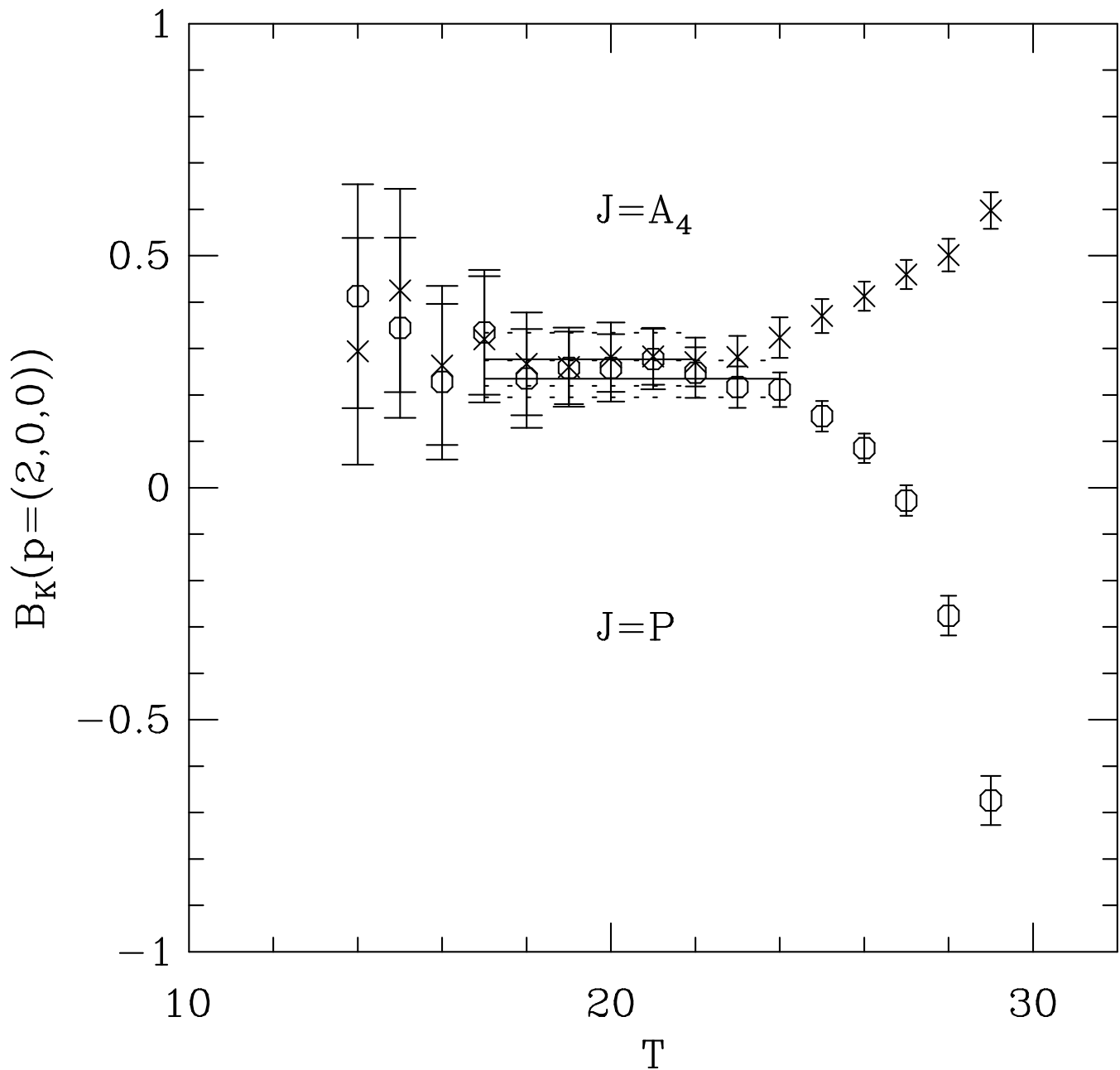}}
\figcaption{Same as Fig.~\protect\ref{f_bk_p0_u2u3} except that 
the data are for momentum transfer $\vec p = (2,0,0)$.
Because the signal in the backward propagating correlator 
(which has the non-zero momentum) dies off for $T \lsim 15$, the 
fit range is asymmetric.}
%\vskip -24pt plus 10pt
\label{f_bk_p2_u2u3}
\end{figure}

A possible source of systematic error is contamination from excited
states. In order to assess the size of this error, we calculate the
ratios with sources having two Lorentz structures: either all are
pseudoscalars, $J \sim P = (\bar\psi_1\gamma_5 \psi_2)$, or all are
axial vectors $J \sim A_4 = (\bar\psi_1\gamma_4\gamma_5\psi_2)$.  The
resulting $B$-parameters should be same in both cases, but we find
that the convergence is from opposite directions, as shown in
Figs.~\ref{f_bk_p0_u2u3} and \ref{f_bk_p2_u2u3}.  If the two
asymptotic results overlap then the effect of contamination is smaller
than the statistical errors.  We find that the two results do overlap,
as illustrated by the Figures, in all channels. The difference between
the two cases increases as the quark mass is decreased, growing to
$\sim 1 \sigma$ for the lightest combination, $U_3U_3$.  In practice,
we take the average of the results from the two sources as our best
estimate.

%% 4/7/96

%% 4/7/96

\subsection{Statistical Errors}
 
We estimate errors using single elimination jackknife.
We first average the data from the forward and
backward propagation of kaons on each configuration. 
Then, within the jackknife loop, we
fit to the ratios for each of the two sources,
and average the results of these fits.

\newsection{Operator Mixing and Perturbative Running}
\label{sec:matching}

In phenomenological applications the matrix elements of operators
must be combined with Wilson coefficients.
These coefficients have been calculated in continuum perturbation
theory in the NDR scheme.
Results are available at a number of scales,
and we take $2 \GeV$ as our standard.
We thus need to relate our results for lattice matrix elements to
those for continuum operators in the NDR scheme at $2 \GeV$.

The general form of the matching formula is
\begin{equation}
\CO^{\rm cont}(\mu)_i = Z(\mu,a)_{ij} \CO^{\rm latt}(a)_j \,,
\end{equation}
where $i$ and $j$ label operators.
$Z$ can be expanded as a power series in $g^2$,
and typically only the $O(g^2)$ term is known.
Rather than use the direct perturbative expansion, however,
it is better to use a renormalization group improved expression.
In Appendix \ref{app:match} we give such an expression for $Z$ 
[Eq.~(\ref{eq:Jimatch2})]
by combining the exact perturbative result of Ji~\cite{Ji} 
with the improved lattice perturbation theory of Lepage and Mackenzie~\cite{Lepage}.
The additional input needed is the continuum two-loop anomalous
dimension matrix.
The improvement is significant numerically when
the difference between continuum scale $\mu=2\GeV$
and the lattice scales $1/a- \pi/a$ becomes substantial,
and when the anomalous dimensions are large. 
The improvement is most important here in the calculations
of $B_{7,8}^{3/2}$.

In practice we use a slightly different improved expression, that from
``horizontal matching'', which is given in Eq.~(\ref{eq:horizontal}).
The numerical difference between the two methods is smaller than
our statistical errors.

We need the matching coefficients for the operators with Dirac structure
$\gamma_\mu L \otimes \gamma_\mu L$ (for $B_K$),
$\gamma_\mu L \otimes \gamma_\mu R$ (for $B_7$ and $B_8$),
and $L \otimes L$ (for $B_S$).
Those for the first two structures have been previously calculated in
Refs. \cite{pertmarti,pertucla}, using the DRED (dimensional reduction)
scheme or a variant thereof.
We have extended these calculations by (i) 
repeating them for all Lorentz scalar four-fermion operators,
including those needed for $B_S$, and (ii) matching to NDR 
in the continuum rather than DRED.
We give the results in Appendix \ref{app:pert}.
We find that using tadpole improved lattice operators
substantially reduces the one-loop corrections~\cite{Lepage,DC95LANL}.
For the local operators we use, tadpole improvement involves
changing the normalization of quark fields by replacing
$\sqrt{2\kappa}$ with
$\sqrt{8\kappa_c}\sqrt{1-3\kappa/4\kappa_c}$.
We make this replacement in our numerical simulations, even though it 
cancels between the numerator and denominator of the $B$ parameters. 

It turns out that the results for the matching coefficients for local
four-fermion operator can be easily generalized from Wilson
fermions to those for any improved fermion action.
One only needs to know the matching coefficients for the
five bilinear operators.  We present the results in Appendix B in
such a way that this generalization can be straightforwardly implemented.

To carry out the matching we need to choose not only the final
continuum scale $\mu$, but also a scale $q^*$ related to the
truncation of perturbation theory (and explained in detail in 
Appendix \ref{app:match}).
For our final results we take $\mu=2\GeV$ and vary $q^*$
in the range $2 \GeV \le q^* \le \pi/a$, taking $1/a$ for our best estimates.
At intermediate stages it is convenient to take $\mu=q^*=1/a$,
which we call the TAD1 scheme. 
This corresponds to doing horizontal matching at $1/a$, 
but no further running in the continuum. This choice 
has the advantage that our results depend very weakly on
the lattice spacing in physical units---the only dependence arises
when we have to interpolate to the physical kaon mass.

Finally, we explain how we determine the $\MSbar$ coupling constant,
which is needed in the matching formula.
We do this following Ref. \cite{Lepage}, by first solving 
\begin{equation}
- \log\Box
= {4\pi\over3} \alpha_V(3.41/a) (1 - 1.185 \alpha_V)
\end{equation}
for $\alpha_V$, and then converting to the $\MSbar$ scheme using
\begin{equation}
\alpha_{\overline{\rm MS}}(3.41 / a) = 
\alpha_V(e^{5/6}\; 3.41 / a) (1 + 2\alpha_V / \pi) \,.
\end{equation}
For the plaquette we use $\Box=0.5937$.
We run the coupling to other scales using the two-loop $\beta-$function.
This results in $\alpha_s(q^*) = 0.2049$, $0.1927$, $0.1523$ and $0.1343$
for $q^*=2\GeV$, $1/a$, $2/a$ and $\pi/a$, respectively.

\newsection{Results for $B_K$}
\label{sec:BKresults}

The kaon $B$-parameter is defined in the continuum by
\begin{equation}
B_K(\mu) = { \vev{K_0| \CQ | \bar{K_0}} \over
	(8/3)\, \vev{K_0| \bar s_a \gamma_\nu \gamma_5 d_a|0}\,
 	      \vev{0| \bar s_b \gamma_\nu \gamma_5 d_b|\bar{K_0}} } 
\,,
\label{eq:bkdef}
\end{equation}
where $\CQ$ is the left-left $\Delta S=2$ operator
\begin{equation}
  \CQ = \bar s_a \gamma_\mu L d_a \ 
	    \bar s_b \gamma_\mu L d_b\,,
\end{equation}
and $L=(1-\gamma_5)$.
We are ultimately interested in the value of $B_K$ when $\CQ$ does not
insert momentum and when the ``kaons'' have physical masses.
The definition Eq. \ref{eq:bkdef} is, however, more general,
and is useful at intermediate stages.
$B_K$ depends on the renormalization scale, $\mu$, because $\CQ$ has
a non-vanishing anomalous dimension unlike the axial current.

To calculate $B_K$, we use the results of Appendices \ref{app:match} 
and \ref{app:pert} to
match $\CQ$ to a linear combination of lattice operators with 
different Lorentz structures.
Thus $B_K$ itself is a linear combination of lattice $B$-parameters,
which we define to be the ratio of the matrix element of the 
four-fermion operators [listed in Eq.~(\ref{eq:Bklstruct})] to 
\begin{equation}
(4/3) \vev{K_0| (\bar s_a \gamma_\nu \gamma_5 d_a)^{\rm lat}|0}
      \vev{0| (\bar s_b \gamma_\nu \gamma_5 d_b)^{\rm lat}|\bar{K_0}} 
\,.
\label{eq:bdenom}
\end{equation}
The four-fermion operators in Eq.~(\ref{eq:Bklstruct}) have
half the Wick contractions of $\CQ$, which is why Eq.~(\ref{eq:bdenom})
contains $4/3$ in contrast to the $8/3$ in Eq.~(\ref{eq:bkdef}).
Note that we do not include perturbative matching factors in these
$B$-parameters, i.e. the operators in the numerator and denominator are
bare lattice operators.
In Table~\ref{t_bku2u3} we give our results for all the $B$-parameters
for the $U_2 U_3$ meson. This is the
lattice meson whose mass lies close to that of the physical kaon.

We also include in Table~\ref{t_bku2u3} results for $B_K$ using one-loop
matching in the TAD1 scheme, i.e. $\mu=q^*=1/a$. 
At tree level, $B_K$ is just the following sum of the $B$-parameters (for 
definitions see Appendix~\ref{app:pert})
\begin{equation}
B_K \sim \CA_s^1+\CA_t^1 + \CV_s^1 + \CV_t^1 +
	 \CA_s^2+\CA_t^2 + \CV_s^2 + \CV_t^2
\,.
\end{equation}
At one-loop, however, all Lorentz structures enter. 
In the result for $B_K$ we include one-loop matching factors 
not only in the numerator but also in the denominator,
i.e. the lattice axial currents in the
denominator have each been multiplied by $Z_A = 1 + \lambda(q^*) c_A$.
To reduce errors, we always form the appropriate linear combination of
operators in the numerator and denominator before fitting.

\begin{table} %1
%\vskip 6pt
$$
\let\ifspace=\iffalse
\def\myskip{\omit&height1.5pt&%
\omit&&%
\omit&&%
\omit&&%
\omit&&%
\omit&&%
\omit&&%
\omit&&%
\omit&&%
\omit&&%
\omit&&%
 &\cr}
\vbox{\hbox{\vbox{
\tabskip=0pt\offinterlineskip
\def\tlr{\noalign{\hrule}}
\def\CO{{\cal O}}
\def\xx{{\langle \bar \psi \psi \rangle}}
\halign {\strut#& \vrule\vrule#\tabskip=2pt&
  \hfil$#$\hfil&\vrule#&
  \hfil$#$\hfil&\vrule#&
  \hfil$#$\hfil&\vrule#&
  \hfil$#$\hfil&\vrule#&
  \hfil$#$\hfil&\vrule#&
  \hfil$#$\hfil&\vrule#&
  \hfil$#$\hfil&\vrule#&
  \hfil$#$\hfil&\vrule#&
  \hfil$#$\hfil&\vrule#&
  \hfil$#$\hfil&\vrule#&
  \hfil$#$\hfil&\vrule#\tabskip=0pt\cr\tlr
\omit&height1.5pt&\multispan{21   }&\cr
\myskip
&& 
&& P[0]P         
&& A[0]A         
&& P[1]P         
&& A[1]A         
&& P[2]P         
&& A[2]A         
&& P[3]P         
&& A[3]A         
&& P[4]P         
&& A[4]A         
  &\cr
\myskip\tlr
\omit&height0.5pt&\multispan{21   }&\cr\tlr
\myskip
&& -{\cal P}^1      
&&  13.9( 1)
&&  13.4( 2)
&&   9.9( 2)
&&   9.6( 2)
&&   8.5( 3)
&&   7.9( 3)
&&   7.5( 4)
&&   6.7( 3)
&&   7.1( 4)
&&   5.5( 4)
  &\cr\ifspace\myskip&&
&&\hbox{  11 --   21}
&&\hbox{  11 --   21}
&&\hbox{  11 --   21}
&&\hbox{  11 --   21}
&&\hbox{  16 --   22}
&&\hbox{  16 --   23}
&&\hbox{  16 --   23}
&&\hbox{  17 --   24}
&&\hbox{  18 --   25}
&&\hbox{  18 --   24}
  &\cr\fi\myskip\tlr
\myskip
&& -{\cal S}^1      
&&   3.36(  4)
&&   3.22(  7)
&&   2.31(  8)
&&   2.26( 10)
&&   2.05( 12)
&&   1.89( 11)
&&   1.79( 17)
&&   1.57( 14)
&&   1.50( 21)
&&   1.20( 17)
  &\cr\ifspace\myskip&&
&&\hbox{  11 --   21}
&&\hbox{  11 --   21}
&&\hbox{  11 --   21}
&&\hbox{  11 --   21}
&&\hbox{  16 --   22}
&&\hbox{  16 --   23}
&&\hbox{  16 --   23}
&&\hbox{  17 --   24}
&&\hbox{  18 --   25}
&&\hbox{  18 --   24}
  &\cr\fi\myskip\tlr
\myskip
&& {\cal V}_s^1    
&&    4.43(  5)
&&    4.20(  9)
&&    2.88(  9)
&&    2.70( 12)
&&    2.41( 13)
&&    2.11( 12)
&&    2.06( 18)
&&    1.66( 15)
&&    1.78( 24)
&&    1.22( 21)
  &\cr\ifspace\myskip&&
&&\hbox{  11 --   21}
&&\hbox{  11 --   21}
&&\hbox{  11 --   21}
&&\hbox{  11 --   21}
&&\hbox{  16 --   22}
&&\hbox{  16 --   23}
&&\hbox{  16 --   23}
&&\hbox{  17 --   24}
&&\hbox{  18 --   25}
&&\hbox{  18 --   24}
  &\cr\fi\myskip\tlr
\myskip
&& {\cal V}_t^1    
&&    1.62(  2)
&&    1.54(  3)
&&    1.09(  3)
&&    1.03(  4)
&&    0.93(  4)
&&    0.84(  4)
&&    0.80(  6)
&&    0.68(  5)
&&    0.72(  8)
&&    0.52(  7)
  &\cr\ifspace\myskip&&
&&\hbox{  11 --   21}
&&\hbox{  11 --   21}
&&\hbox{  11 --   21}
&&\hbox{  11 --   21}
&&\hbox{  16 --   22}
&&\hbox{  16 --   23}
&&\hbox{  16 --   23}
&&\hbox{  17 --   24}
&&\hbox{  18 --   25}
&&\hbox{  18 --   24}
  &\cr\fi\myskip\tlr
\myskip
&& -{\cal A}_s^1    
&&   3.46(  4)
&&   3.30(  6)
&&   2.34(  6)
&&   2.21(  8)
&&   1.95(  8)
&&   1.73(  8)
&&   1.68( 11)
&&   1.38(  9)
&&   1.51( 15)
&&   1.01( 13)
  &\cr\ifspace\myskip&&
&&\hbox{  11 --   21}
&&\hbox{  11 --   21}
&&\hbox{  11 --   21}
&&\hbox{  11 --   21}
&&\hbox{  16 --   22}
&&\hbox{  16 --   23}
&&\hbox{  16 --   23}
&&\hbox{  17 --   24}
&&\hbox{  18 --   25}
&&\hbox{  18 --   24}
  &\cr\fi\myskip\tlr
\myskip
&& -{\cal A}_t^1    
&&   0.86(  1)
&&   0.81(  2)
&&   0.49(  2)
&&   0.45(  2)
&&   0.37(  3)
&&   0.30(  2)
&&   0.29(  4)
&&   0.19(  3)
&&   0.22(  5)
&&   0.08(  5)
  &\cr\ifspace\myskip&&
&&\hbox{  11 --   21}
&&\hbox{  11 --   21}
&&\hbox{  11 --   21}
&&\hbox{  11 --   21}
&&\hbox{  16 --   22}
&&\hbox{  16 --   23}
&&\hbox{  16 --   23}
&&\hbox{  17 --   24}
&&\hbox{  18 --   25}
&&\hbox{  18 --   24}
  &\cr\fi\myskip\tlr
\myskip
&& {\cal T}_s^1    
&&   15.2( 2)
&&   14.6( 2)
&&   10.5( 3)
&&   10.2( 4)
&&    9.0( 4)
&&    8.4( 4)
&&    7.9( 6)
&&    7.0( 5)
&&    7.1( 7)
&&    5.7( 6)
  &\cr\ifspace\myskip&&
&&\hbox{  11 --   21}
&&\hbox{  11 --   21}
&&\hbox{  11 --   21}
&&\hbox{  11 --   21}
&&\hbox{  16 --   22}
&&\hbox{  16 --   23}
&&\hbox{  16 --   23}
&&\hbox{  17 --   24}
&&\hbox{  18 --   25}
&&\hbox{  18 --   24}
  &\cr\fi\myskip\tlr
\myskip
&& {\cal T}_t^1    
&&   15.3( 2)
&&   14.7( 2)
&&   10.7( 3)
&&   10.4( 4)
&&    9.2( 4)
&&    8.5( 4)
&&    8.0( 5)
&&    7.1( 5)
&&    7.3( 7)
&&    5.7( 6)
  &\cr\ifspace\myskip&&
&&\hbox{  11 --   21}
&&\hbox{  11 --   21}
&&\hbox{  11 --   21}
&&\hbox{  11 --   21}
&&\hbox{  16 --   22}
&&\hbox{  16 --   23}
&&\hbox{  16 --   23}
&&\hbox{  17 --   24}
&&\hbox{  18 --   25}
&&\hbox{  18 --   24}
  &\cr\fi\myskip\tlr
\myskip
&& -{\cal P}^2      
&&  37.2( 3)
&&  36.0( 5)
&&  26.7( 5)
&&  26.0( 6)
&&  22.9( 7)
&&  21.3( 7)
&&  20.3( 9)
&&  18.0( 8)
&&  19.4(11)
&&  15.1(10)
  &\cr\ifspace\myskip&&
&&\hbox{  11 --   21}
&&\hbox{  11 --   21}
&&\hbox{  11 --   21}
&&\hbox{  11 --   21}
&&\hbox{  16 --   22}
&&\hbox{  16 --   23}
&&\hbox{  16 --   23}
&&\hbox{  17 --   24}
&&\hbox{  18 --   25}
&&\hbox{  18 --   24}
  &\cr\fi\myskip\tlr
\myskip
&& -{\cal S}^2      
&&   4.78(  8)
&&   4.59( 12)
&&   3.30( 15)
&&   3.27( 20)
&&   3.06( 23)
&&   2.79( 20)
&&   2.66( 32)
&&   2.30( 24)
&&   2.13( 36)
&&   1.76( 30)
  &\cr\ifspace\myskip&&
&&\hbox{  11 --   21}
&&\hbox{  11 --   21}
&&\hbox{  11 --   21}
&&\hbox{  11 --   21}
&&\hbox{  16 --   22}
&&\hbox{  16 --   23}
&&\hbox{  16 --   23}
&&\hbox{  17 --   24}
&&\hbox{  18 --   25}
&&\hbox{  18 --   24}
  &\cr\fi\myskip\tlr
\myskip
&& {\cal V}_s^2    
&&    0.40(  1)
&&    0.37(  1)
&&    0.25(  1)
&&    0.24(  2)
&&    0.19(  1)
&&    0.17(  2)
&&    0.16(  2)
&&    0.12(  2)
&&    0.14(  2)
&&    0.05(  3)
  &\cr\ifspace\myskip&&
&&\hbox{  11 --   21}
&&\hbox{  11 --   21}
&&\hbox{  11 --   21}
&&\hbox{  11 --   21}
&&\hbox{  16 --   22}
&&\hbox{  16 --   23}
&&\hbox{  16 --   23}
&&\hbox{  17 --   24}
&&\hbox{  18 --   25}
&&\hbox{  18 --   24}
  &\cr\fi\myskip\tlr
\myskip
&& {\cal V}_t^2    
&&    0.15(  0)
&&    0.14(  0)
&&    0.10(  0)
&&    0.09(  1)
&&    0.08(  1)
&&    0.08(  1)
&&    0.08(  1)
&&    0.06(  1)
&&    0.06(  1)
&&    0.04(  1)
  &\cr\ifspace\myskip&&
&&\hbox{  11 --   21}
&&\hbox{  11 --   21}
&&\hbox{  11 --   21}
&&\hbox{  11 --   21}
&&\hbox{  16 --   22}
&&\hbox{  16 --   23}
&&\hbox{  16 --   23}
&&\hbox{  17 --   24}
&&\hbox{  18 --   25}
&&\hbox{  18 --   24}
  &\cr\fi\myskip\tlr
\myskip
&& -{\cal A}_s^2    
&&   2.87(  3)
&&   2.76(  4)
&&   1.98(  4)
&&   1.88(  6)
&&   1.66(  6)
&&   1.54(  7)
&&   1.44(  9)
&&   1.27(  8)
&&   1.35( 12)
&&   1.05( 11)
  &\cr\ifspace\myskip&&
&&\hbox{  11 --   21}
&&\hbox{  11 --   21}
&&\hbox{  11 --   21}
&&\hbox{  11 --   21}
&&\hbox{  16 --   22}
&&\hbox{  16 --   23}
&&\hbox{  16 --   23}
&&\hbox{  17 --   24}
&&\hbox{  18 --   25}
&&\hbox{  18 --   24}
  &\cr\fi\myskip\tlr
\myskip
&& {\cal A}_t^2    
&&   -0.17(  1)
&&   -0.14(  1)
&&    0.12(  1)
&&    0.14(  2)
&&    0.20(  3)
&&    0.25(  3)
&&    0.25(  4)
&&    0.33(  4)
&&    0.33(  5)
&&    0.39(  5)
  &\cr\ifspace\myskip&&
&&\hbox{  11 --   21}
&&\hbox{  11 --   21}
&&\hbox{  11 --   21}
&&\hbox{  11 --   21}
&&\hbox{  16 --   22}
&&\hbox{  16 --   23}
&&\hbox{  16 --   23}
&&\hbox{  17 --   24}
&&\hbox{  18 --   25}
&&\hbox{  18 --   24}
  &\cr\fi\myskip\tlr
\myskip
&& {\cal T}_s^2    
&&    0.54(  1)
&&    0.51(  1)
&&    0.35(  1)
&&    0.34(  1)
&&    0.29(  2)
&&    0.27(  2)
&&    0.25(  2)
&&    0.23(  2)
&&    0.21(  3)
&&    0.16(  3)
  &\cr\ifspace\myskip&&
&&\hbox{  11 --   21}
&&\hbox{  11 --   21}
&&\hbox{  11 --   21}
&&\hbox{  11 --   21}
&&\hbox{  16 --   22}
&&\hbox{  16 --   23}
&&\hbox{  16 --   23}
&&\hbox{  17 --   24}
&&\hbox{  18 --   25}
&&\hbox{  18 --   24}
  &\cr\fi\myskip\tlr
\myskip
&& {\cal T}_t^2    
&&    0.55(  1)
&&    0.52(  1)
&&    0.36(  1)
&&    0.35(  2)
&&    0.30(  2)
&&    0.28(  2)
&&    0.27(  2)
&&    0.24(  2)
&&    0.22(  3)
&&    0.17(  3)
  &\cr\ifspace\myskip&&
&&\hbox{  11 --   21}
&&\hbox{  11 --   21}
&&\hbox{  11 --   21}
&&\hbox{  11 --   21}
&&\hbox{  16 --   22}
&&\hbox{  16 --   23}
&&\hbox{  16 --   23}
&&\hbox{  17 --   24}
&&\hbox{  18 --   25}
&&\hbox{  18 --   24}
  &\cr\fi\myskip\tlr
\myskip
&& {B_K} 
&&   -0.30(  1)
&&   -0.28(  1)
&&   -0.03(  1)
&&   -0.01(  2)
&&    0.09(  2)
&&    0.14(  3)
&&    0.19(  3)
&&    0.23(  3)
&&    0.23(  4)
&&    0.28(  6)
  &\cr\ifspace\myskip&&
&&\hbox{  11 --   21}
&&\hbox{  11 --   21}
&&\hbox{  11 --   21}
&&\hbox{  11 --   21}
&&\hbox{  14 --   21}
&&\hbox{  17 --   23}
&&\hbox{  15 --   22}
&&\hbox{  18 --   23}
&&\hbox{  17 --   24}
&&\hbox{  17 --   22}
  &\cr\fi\myskip\tlr
\myskip
&& {B}_7     
&&    0.60(  0)
&&    0.60(  1)
&&    0.61(  1)
&&    0.60(  1)
&&    0.60(  1)
&&    0.61(  2)
&&    0.60(  2)
&&    0.62(  2)
&&    0.64(  2)
&&    0.62(  3)
  &\cr\ifspace\myskip&&
&&\hbox{  11 --   21}
&&\hbox{  11 --   21}
&&\hbox{  11 --   21}
&&\hbox{  11 --   21}
&&\hbox{  16 --   22}
&&\hbox{  16 --   23}
&&\hbox{  16 --   23}
&&\hbox{  17 --   24}
&&\hbox{  18 --   25}
&&\hbox{  18 --   24}
  &\cr\fi\myskip\tlr
\myskip
&& {B}_8     
&&    0.82(  0)
&&    0.82(  1)
&&    0.82(  1)
&&    0.82(  1)
&&    0.82(  2)
&&    0.82(  2)
&&    0.81(  3)
&&    0.84(  3)
&&    0.86(  3)
&&    0.84(  4)
  &\cr\ifspace\myskip&&
&&\hbox{  11 --   21}
&&\hbox{  11 --   21}
&&\hbox{  11 --   21}
&&\hbox{  11 --   21}
&&\hbox{  16 --   22}
&&\hbox{  16 --   23}
&&\hbox{  16 --   23}
&&\hbox{  17 --   24}
&&\hbox{  18 --   25}
&&\hbox{  18 --   24}
  &\cr\fi\myskip\tlr
\myskip
&& {B}_4^+
&&    0.59(  0)
&&    0.59(  1)
&&    0.60(  1)
&&    0.59(  1)
&&    0.61(  2)
&&    0.61(  2)
&&    0.61(  2)
&&    0.61(  2)
&&    0.61(  3)
&&    0.60(  3)
   &\cr\ifspace\myskip&&
 &&\hbox{  12 --   20}
 &&\hbox{  12 --   20}
 &&\hbox{  12 --   20}
 &&\hbox{  12 --   20}
 &&\hbox{  13 --   21}
 &&\hbox{  13 --   21}
 &&\hbox{  15 --   23}
 &&\hbox{  15 --   23}
 &&\hbox{  18 --   24}
 &&\hbox{  18 --   24}
  &\cr\fi\myskip\tlr
\myskip
&& {B}_5^+
&&    0.78(  0)
&&    0.78(  1)
&&    0.79(  1)
&&    0.78(  2)
&&    0.81(  2)
&&    0.80(  3)
&&    0.80(  3)
&&    0.82(  3)
&&    0.81(  4)
&&    0.80(  4)
   &\cr\ifspace\myskip&&
 &&\hbox{  12 --   20}
 &&\hbox{  12 --   20}
 &&\hbox{  12 --   20}
 &&\hbox{  12 --   20}
 &&\hbox{  13 --   21}
 &&\hbox{  13 --   21}
 &&\hbox{  15 --   23}
 &&\hbox{  15 --   23}
 &&\hbox{  18 --   24}
 &&\hbox{  18 --   24}
  &\cr\fi\myskip\tlr
\cr}}}}$$

%\vskip -12pt plus 10pt
\caption{ $B$-parameters (defined in the text) for individual
operators without any renormalization factors. The results for $B_K$,
$B_7^{3/2}$, $B_8^{3/2}$, $B_4^+$ and $B_5^+$ are in the TAD1
scheme.  All data are for the mass combination $U_2U_3$.  The data are
shown separately for the two types of source $J$ and for the five
momentum transfers. The columns are labeled by $J[p^2]J$.\looseness=-1
}
\label{t_bku2u3}
\end{table}
%% data from Table 14 in Bk.tex (round the data to fit)
%% 4/8/96

The results in the Table indicate the relative size of the matrix elements
of the contributing operators. The largest are the those of the 
pseudoscalars and tensors. The results using the two sources ($J=A_4$ and $P$)
agree within errors in most cases. Our best estimate, as explained above,
is obtained by averaging the two results.
The errors increase as the momentum inserted by the operator increases,
but we retain good statistical control for all five momenta.
This allows us to see clearly that most of the
$B$-parameters have substantial momentum dependence.

\begin{table} %2
%\vskip 6pt
\newcommand\0{\hphantom{0}}
\newcommand\ce[1]{\multicolumn{#1}{c}}
\setlength{\tabcolsep}{6pt}
\begin{tabular}{lrrrrrr}
\hline
% &\ce6{$B_K$}\cr 
&\ce1{$(0,0,0)$}&\ce1{$(1,0,0)$}&\ce1{$(1,1,0)$}&\ce1{$(1,1,1)$}&\ce1{$(2,0,0)$}&\ce1{subtracted}\cr
\hline
$CC    $&$ 0.920(10)$&$ 0.921(11)$&$ 0.924(14)$&$ 0.925(15)$&$ 0.927(16)$&$           $\cr
$CS    $&$ 0.834(09)$&$ 0.838(11)$&$ 0.841(14)$&$ 0.844(16)$&$ 0.847(19)$&$           $\cr
$CU_1  $&$ 0.808(09)$&$ 0.811(12)$&$ 0.813(16)$&$ 0.815(19)$&$ 0.817(22)$&$           $\cr
$CU_2  $&$ 0.796(11)$&$ 0.799(14)$&$ 0.797(20)$&$ 0.799(23)$&$ 0.799(27)$&$           $\cr
$CU_3  $&$ 0.792(13)$&$ 0.795(17)$&$ 0.789(25)$&$ 0.792(29)$&$ 0.787(36)$&$           $\cr
$SS    $&$ 0.558(04)$&$ 0.586(07)$&$ 0.605(10)$&$ 0.623(13)$&$ 0.638(16)$&$ 0.746(216)$\cr
$SU_1  $&$ 0.444(04)$&$ 0.488(07)$&$ 0.518(11)$&$ 0.545(15)$&$ 0.567(19)$&$ 0.730(165)$\cr
$SU_2  $&$ 0.382(05)$&$ 0.437(08)$&$ 0.473(12)$&$ 0.506(17)$&$ 0.532(22)$&$ 0.716(146)$\cr
$SU_3  $&$ 0.338(05)$&$ 0.402(09)$&$ 0.442(15)$&$ 0.478(20)$&$ 0.508(26)$&$ 0.704(135)$\cr
$U_1U_1$&$ 0.244(04)$&$ 0.332(08)$&$ 0.387(13)$&$ 0.433(17)$&$ 0.466(23)$&$ 0.716(115)$\cr
$U_1U_2$&$ 0.112(05)$&$ 0.237(09)$&$ 0.310(15)$&$ 0.369(21)$&$ 0.408(28)$&$ 0.703(098)$\cr
$U_1U_3$&$ 0.000(07)$&$ 0.161(10)$&$ 0.248(18)$&$ 0.318(25)$&$ 0.363(34)$&$ 0.687(088)$\cr
$U_2U_2$&$-0.091(07)$&$ 0.103(11)$&$ 0.204(18)$&$ 0.283(26)$&$ 0.330(38)$&$ 0.688(082)$\cr
$U_2U_3$&$-0.287(10)$&$-0.017(14)$&$ 0.112(22)$&$ 0.212(33)$&$ 0.256(49)$&$ 0.670(075)$\cr
$U_3U_3$&$-0.597(16)$&$-0.191(21)$&$-0.021(31)$&$ 0.112(45)$&$ 0.140(68)$&$ 0.643(074)$\cr
\hline
\end{tabular}

%\vskip -12pt plus 10pt
\caption{Lattice results for the $B$-parameter of the $\Delta S = 2$
operator $\CQ$ in the TAD1 scheme.  The momentum inserted by the
operator is indicated at the top of the columns, and the rows label
the meson's quark combination. Results after momentum subtraction (see
text) are given for the ten lightest mesons.}%
\label{t_bkbufs}
\end{table}
%% data from Table in auxtabs.tex 
%% 11/20/95

In Table \ref{t_bkbufs} we give our results for $B_K$ for
all mass combinations.
To extract the physical value of $B_K$ we must account for
the breaking of chiral symmetry by Wilson fermions.
The general form of the chiral expansion of the
matrix elements of four-fermion operators is
\looseness-1
\begin{equation}
{\left\langle \overline{K^0}(p_f) 
\right| {\cal O} \left| K^0(p_i) \right\rangle 
\over   (8/3) f_{K,{\rm phys}}^2}  = 
\alpha + \beta m_K^2 + \gamma \, p_i\cdot p_f + \delta_1 m_K^4 + 
\delta_2 m_K^2 p_i\cdot p_f + \delta_3 (p_i\cdot p_f )^2 + \ldots .
\label{eq:Bkcpt}
\end{equation}
%% \begin{eqnarray}
%% {\left\langle \overline{K^0}(p_f) 
%% \right| {\cal O} \left| K^0(p_i) \right\rangle 
%% \over   (8/3) f_{K,{\rm phys}}^2}  ={}&
%% \alpha + \beta m_K^2 + \gamma \, p_i\cdot p_f + \delta_1 m_K^4 + 
%% \aftergroup\hfill \nonumber\\
%% {}& \delta_2 m_K^2 p_i\cdot p_f + \delta_3 (p_i\cdot p_f )^2 + \ldots .
%% \label{eq:Bkcpt}
%% \end{eqnarray}
where $f_{K,{\rm phys}}$ is the physical kaon decay constant 
(in lattice units).
We are ignoring chiral logarithms and terms proportional to $(m_s-m_d)^2$.
The former are difficult to distinguish numerically from the
terms we include, while the latter we expect to be small, especially
for the range of quark masses studied.  We are also assuming that we
do not need to include possible $p^4$ terms which violate Lorentz symmetry.  

For most operators we expect all terms in Eq.~(\ref{eq:Bkcpt}) 
to contribute. For $\CO=\CQ^{\rm cont}$, however,
chiral perturbation theory predicts that the matrix element vanishes when 
$p_i\cdot p_f=0$, requiring $\alpha$, $\beta$ and $\delta_1$ to be zero.  
This means that for $B_K$, one expects
\begin{equation}
B_K = 
{\vev{K^0|\CQ^{\rm cont}|\bar{K^0}}  \over   (8/3) f_K^2 p_i \cdot p_f} 
% = {f_{K,{\rm phys}}^2} \over f_K^2 p_i \cdot p_f}
% {\vev{K^0|\CQ^{\rm cont}|\bar{K^0}}  \over   (8/3) f_{K,{\rm phys}}^2} 
= {{f_{K,{\rm phys}}^2} \over f_K^2} 
( \gamma + \delta_2 m_K^2 +\delta_3 p_i\cdot p_f) \,.
\end{equation}
With Wilson fermions, however, chiral symmetry is broken explicitly.
This would not be a problem if one
could match with the continuum operator using a nonperturbative
method, and if one could extrapolate to the continuum limit.  We can
do neither of these, and so expect non-zero values of $\alpha$,
$\beta$ and $\delta_1$ proportional to $a g^2$ and to $g^4$.  There
will also be artifacts of this order contained in the other
coefficients in the chiral expansion.

%We give our results for $B_K(\mu=q^*=1/a)$ for all quark mass
%combinations in Table \ref{t_bkbufs}. The data show
%that $\alpha < 0$---and $B_K$ does not 
%tend to a constant at small quark masses and momenta.

\begin{figure} %4
\hbox{\epsfxsize=\hsize\epsfbox{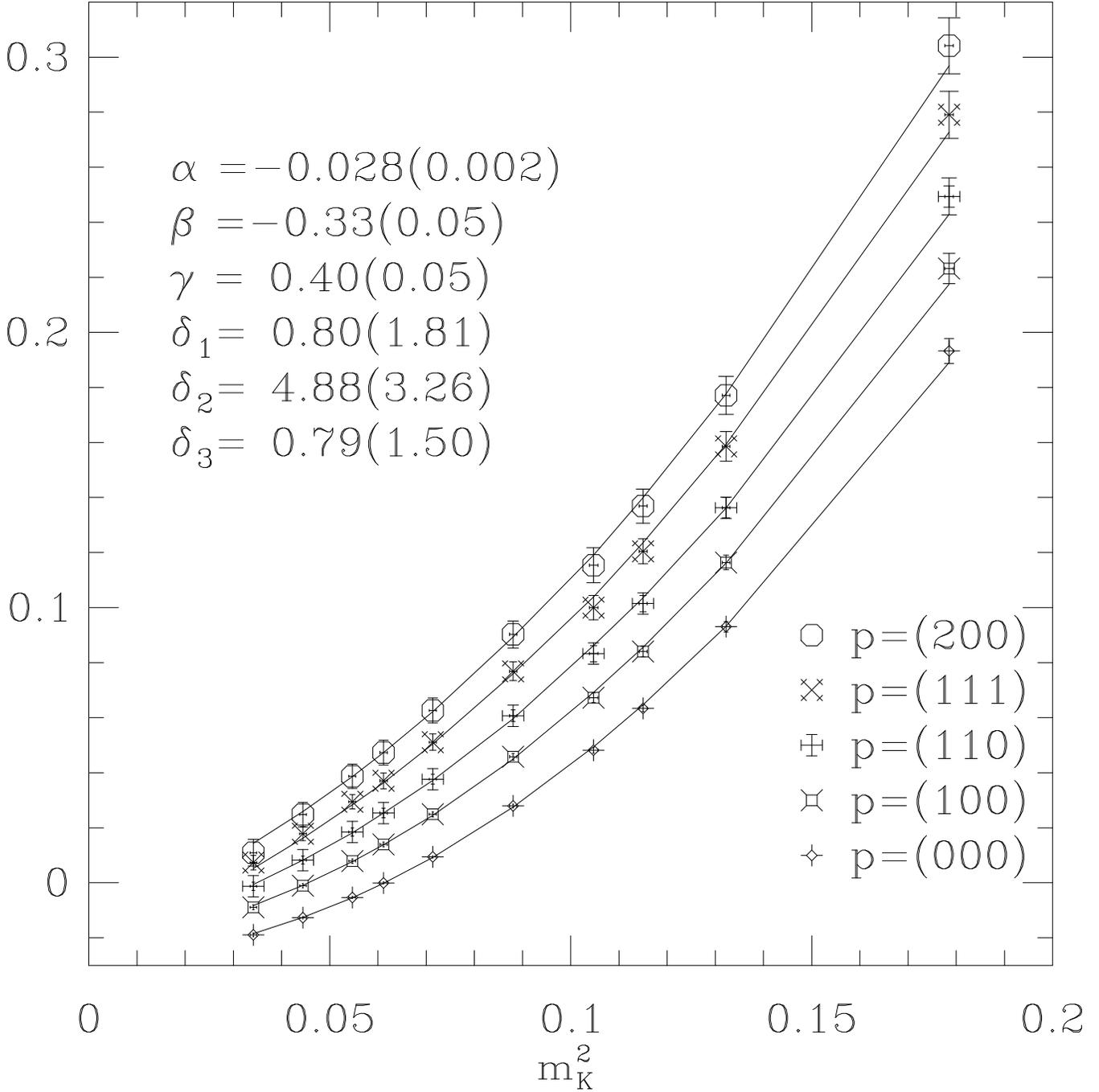}}
\figcaption{Results for the matrix element
$\CQ^{\rm cont}/[8 f_{K,{\rm phys}}^2/3]$ (in TAD1 scheme),
together with the chiral fit. 
The physical kaon corresponds to
$m_{K,{\rm phys}}^2=0.046$ in lattice units.
}
%\vskip -24pt plus 10pt
\label{f_bkcpt}
\end{figure}
%% Figure 21 in mefitplots.ax (to update replace data and coefficients)
%% 4/08/96

To show that this is indeed a problem, we plot our results
for ${\vev{K^0|\CQ^{\rm cont}|\bar{K^0}}/[8 f_{K,{\rm phys}}^2}/3]$
in Fig.~\ref{f_bkcpt}.
The figure also
shows the result of fitting the data with the six-parameter form of
Eq.~(\ref{eq:Bkcpt}).  A good fit is obtained, with non-zero values
for the artifacts $\alpha$ and $\beta$. The curvature in the data
indicates the presence of $p^4$ terms, but the values of the
individual $\delta_i$ are poorly determined.

\begin{figure} %5
\hbox{\epsfxsize=\hsize\epsfbox{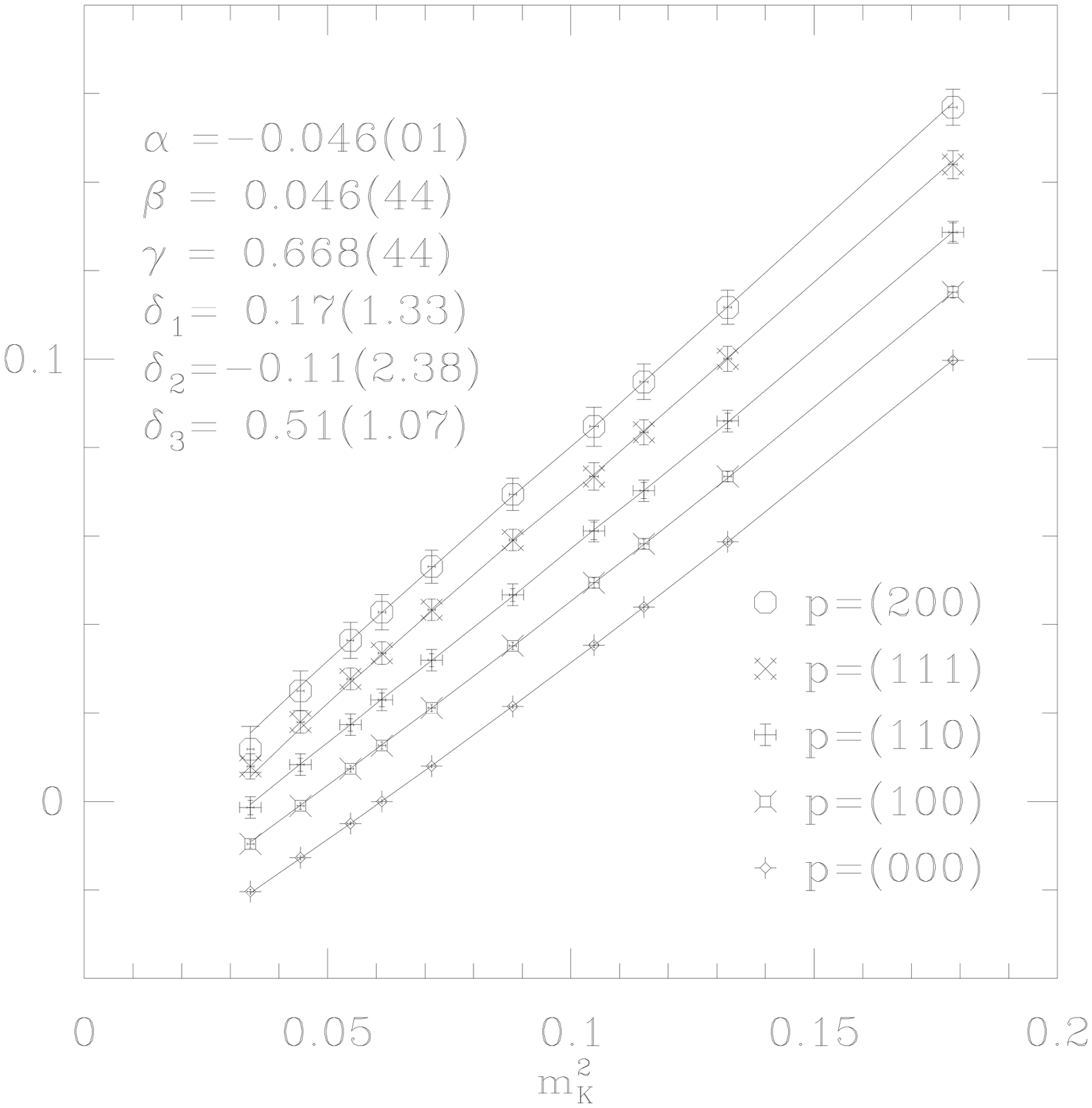}}
\figcaption{Results for matrix element
$\CQ^{\rm cont}/[8 f_{K}^2/3]$,
together with the results of a chiral fit.
}
%\vskip -24pt plus 10pt
\label{f_APEcpt}
\end{figure}

The curvature is drastically reduced if we consider the ratio 
\begin{equation}
{\vev{K^0|\CQ|\bar{K^0}}  \over   (8/3) f_K^2}
= B_K p_i \cdot p_f
\,,
\end{equation}
as shown in Fig.~\ref{f_APEcpt}.
Here $f_K$ is the lattice result for
the decay constant for the corresponding meson \cite{DC95LANL}.
This simplification has been noted previously
by the APE group \cite{Bk95APE}.
The figure also shows a fit to the same form as on the r.h.s. of
Eq. \ref{eq:Bkcpt}.
Not only are the higher order terms smaller 
(although they remain poorly determined), 
but also $\beta$ is much reduced.
Fitting to this ratio leads to smaller errors in the final result,
and we use it in all of the following analysis.

\begin{figure} 
\hbox{\epsfxsize=\hsize\epsfbox{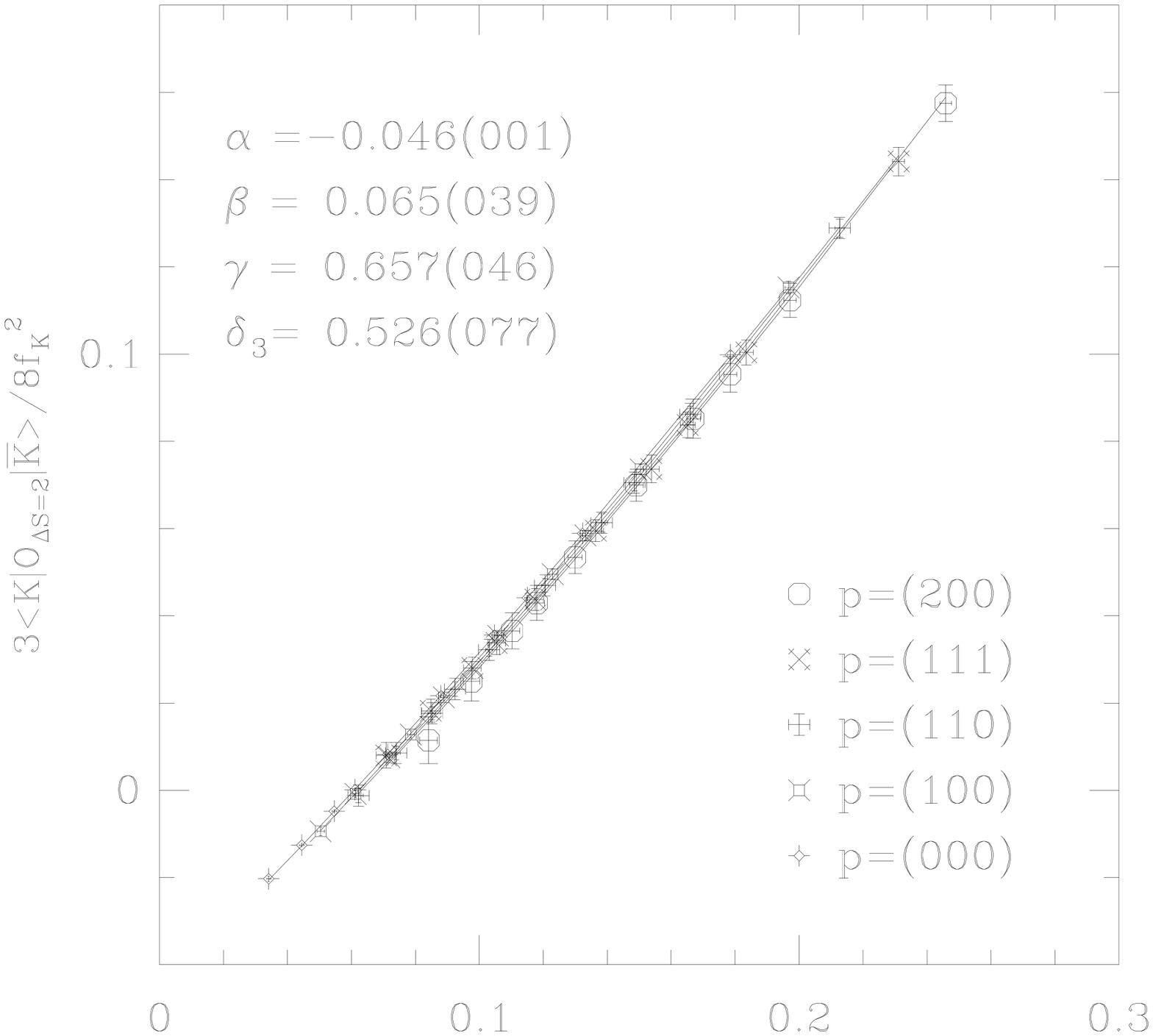}}
\figcaption{Results for matrix element of $\CQ^{\rm cont}/[8 f_{K}^2/3]$,
together with the results of a chiral fit excluding $\delta_1$
and $\delta_2$. The data are plotted as a function of $p_i \cdot p_f$.}
%\vskip -24pt plus 10pt
\label{f_bestBkfit}
\end{figure}
%% extracted from mefitplots.ax -- FIT_NUMBER 9A

We can reduce the uncertainty in the higher order terms
by noting that the matrix element of $\CQ^{\rm cont}/[8 f_{K}^2/3]$ 
is to good approximation a function only of $p_i \cdot p_f$,
as shown in Fig.~\ref{f_bestBkfit}.
This feature was also found in Ref.~\cite{Bk95APE}.
This suggests fitting with $\delta_1=\delta_2=0$,
and the result of such a fit is shown in the figure.
Thus the dominant artifact is $\alpha$, with a small contribution 
from $\beta$.
that leads to the observed deviations from a single curve.  This fit
leads to the smallest errors in $B_K$, and we take it as our standard.

In order to extract a value for $B_K$ we use a procedure similar
to that we advocated in Ref. \cite{Bk93LANL}, except that we 
now have better control over the higher order terms.
Having obtained a fit to the data, we simply discard the terms
which we know to be artifacts. Applying this to the fit shown
in Fig. ~\ref{f_bestBkfit} gives
\begin{equation}
B_K(\mu=q^*=1/a) 
\approx \gamma + (\delta_2 + \delta_3) M_{K,{\rm phys}}^2 
= 0.68(4)\,.
\label{eq:Bkmomfit}
\end{equation}
We obtain consistent results, with larger errors, from the fits of
Fig.~\ref{f_APEcpt} [$0.68(6)$] and
Fig.~\ref{f_bkcpt} [$0.65(10)$].
The subtraction procedure has, however, increased the errors substantially
compared to those in the values for individual mass combinations.
For example, the error for $U_2U_3$ in $B_K$ is 0.01 
(see Table~\ref{t_bku2u3}).

To give an idea of the importance of the $p^4$ terms we have
done a variety of alternate fits, all to the ratio $\CQ/[8 f_K^2/3]$.
Most interesting are those using only the three lowest
order terms ($\alpha$, $\beta$ and $\gamma$).
Fitting to the results from the lightest four mesons alone gives
$B_K=0.73(4)$, while fitting to the lightest six gives $0.74(4)$.
These are higher than our preferred number, but the difference
is only slightly more than $1\sigma$.
By contrast, if we fit to the combinations $U_1 U_1$, $S U_i$ and $SS$ 
(the heaviest five of the mesons in Figs.~\ref{f_bkcpt} and \ref{f_APEcpt}), 
we find a result, $0.80(4)$, significantly higher than our preferred number.
Previous work by other groups with Wilson fermions\cite{Bk95APE,sonilat95},
including our own \cite{Bk93LANL},
used three-parameter chiral fits to mesons in this mass range. 
Our results indicate that the neglect of higher order terms might 
lead to 10-20\% systematic errors in their results.

We have checked our result by repeating the analysis using a second method.
Returning to the original definition of the chiral expansion,
Eq. (\ref{eq:Bkcpt}), we note that
contributions from the artifacts $\alpha, \beta, \delta_1$ can be canceled
by combining pairs of points at different momentum transfers:
\begin{equation}
{E_1 B_K(q_1) - E_2 B_K(q_2) \over (E_1-E_2)}  = 
\gamma + \delta_2 M^2 + \delta_3 M (E_1+E_2).  
\label{eq:Bkmomsub}
\end{equation}
Here $E_i$ is the energy of the lattice kaon with momentum $\vec q_i$.
The r.h.s. of Eq.~(\ref{eq:Bkmomsub}) differs from the form for $B_K$,
Eq.~(\ref{eq:Bkmomfit}), by $\delta_3 M (E_1+E_2-M)$.
We use the value of $\delta_3$ extracted from the fit in Fig.~\ref{f_APEcpt}
to correct for this.
The results of this analysis for the ten light mass combinations 
are given in the sixth column of Table~\ref{t_bkbufs}. 
We have averaged the results from $[q_{1},q_2]=[(100),(000)]$
and $[q_{1},q_2]=[(110),(000)]$, the channels with the best signal.
Extrapolating/interpolating to $M_K$, we find $B_K = 0.66(9)$,
in good agreement with our earlier result.

We thus are confident that our results do not have significant errors
due to yet higher order terms (of $O(p^6)$) in the chiral expansion.
What is more difficult to gauge, however, is the size of the lattice
artifacts in the chiral coefficients which we keep, 
i.e. $\gamma$, $\delta_2$ and $\delta_3$.
These artifacts are due both to discretization errors and to higher order
terms in the perturbative matching coefficients.
Recently the APE group has studied $B_K$ using Wilson fermions and 
the improved ``clover'' action, which removes errors of $O(a g^2)$ 
\cite{Bk95APE}.
Their result suggests that the largest source of error in our calculation
is from the matching coefficients rather than discretization.

Assuming that truncated perturbation theory is the problem, we can
estimate the size of the associated systematic errors using a method
adapted from that of Bernard and Soni \cite{Bk89SONI}.  As
explained in Appendix \ref{app:pert}, $\CQ$ matches with a linear
combination of five lattice operators: itself ($2\CO^+_1$) and the
four other Fierz self-conjugate operators $\CO^+_{2-5}$.  The idea is
to separately adjust the matching coefficients of $\CO^+_{2-5}$
relative to that of $\CO^+_1$ so as to remove the artifacts $\alpha$,
$\beta$ and $\delta_1$.  The change in matching coefficients would
lead to a change in the value of $B_K$.  Since there are four
operators to adjust but only three coefficients to fix at $O(p^4)$, we
need to make additional assumptions to carry out this procedure.  One
approach might be to consider three operators at a time, and use the
spread in the final results as an indicator of the systematic error.

In practice, we have only been able to carry out this procedure 
semi-quantitatively because our statistical errors are too large.
In table~\ref{t_yfits} we show the results of chiral fits to matrix
elements of various operators keeping only $\alpha$, $\gamma$, and
$\delta_3$:
\begin{enumerate}
\item
The perturbatively corrected $\Delta S=2$ operator $\CQ$ (in the TAD1 scheme)
divided by  $8 f_K^2/3$. This is the matrix element relevant for $B_K$,
the results for which are shown in Fig.~\ref{f_bestBkfit}.
\item
Four linear combinations of the bare lattice operators 
$\CO^+_{2-5}$, each divided by $8 f_K^2/3$.
\end{enumerate}
The three parameter fit requires that the data lie on a single curve
when plotted against $p_i \cdot p_f$, which is not a bad representation
of the $\CQ$ data, as shown in Fig.~\ref{f_bestBkfit}.
The fit to $\CO^+_{2-5}$ is better still.
We do not introduce the fourth parameter $\beta$, because, although this
improves the fit for $\CQ$, it makes little difference for $\CO^+_{2-5}$,
and increases the errors in the fit parameters.

\begin{table} %3
\renewcommand{\arraystretch}{1.2}
\begin{center}
\begin{tabular}{lccc}
\hline
Operator 		& $\alpha$ 	& $\gamma$	& $\delta_3$ \\
\hline
$\CQ$ 			& -0.0457(14)	& 0.713(20)	& 0.52(7) \\
$\CO^+_2+\CO^+_3$ 	& 1.79(5)	& 1.1(5)	&-1.5(1.4) \\
$\CO^+_2-\CO^+_3$ 	&-3.36(9)	& -4.8(9)	& 2.9(2.9) \\
$2(\CO^+_4+\CO^+_5)$ 	&-1.29(4)	& -1.8(4)	& 0.3(1.2) \\
$2(\CO^+_4-\CO^+_5)/3$ 	&-0.71(2)	& -0.8(2)	& 0.0(6) \\
\hline
\end{tabular}
\caption{Results (in lattice units)
of three parameter chiral fits to various operators.}
\label{t_yfits}
\end{center}
\end{table}
%% Results from TAble for Fit 9

Since our data is reasonably represented with only the single artifact
$\alpha$, and since the four ``subtraction operators'' can be
similarly represented, our subtraction procedure is as follows.
Take each of the four operators listed in the table in turn, 
add them to $\CQ$ so as to cancel $\alpha$, and use the change
in $\gamma + \delta_3 m_{K,\rm phys}^2$ as the estimate of the 
possible shift in $B_K$. In fact we ignore
the contribution of $\delta_3$, since it is both small and uncertain.
The shifts in $B_K$ are then 0.03, 0.07, 0.06, and 0.05 for the
four subtraction operators listed in the Table. 
Thus we take $0.05 \pm 0.05$ for our estimate of this shift, the generous 
error accounting for the uncertainty in this subtraction method.
Thus our final estimate of $B_K$ is 
\begin{equation}
B_K(\mu=q^*=1/a) = 0.73 \pm 0.04 ({\rm stat}) \pm 0.05 ({\rm subt}) \,.
\end{equation}
Running in the continuum to $\mu=2\,$GeV increases the result slightly
\begin{equation}
B_K(\mu=2 \GeV,q^*=1/a) = 0.74 \pm 0.04 ({\rm stat}) \pm 0.05 ({\rm subt}) \,.
\end{equation}

\begin{figure} %10
\figcaption{Three parameter fit to the matrix elements of $\CP^2/[8 f_K^2/3]$.
The data for $\CP^1$ is similar and the VSA relation $\CP^2=3\CP^1$
is approximately valid.}
\hbox{\epsfxsize=\hsize\epsfbox{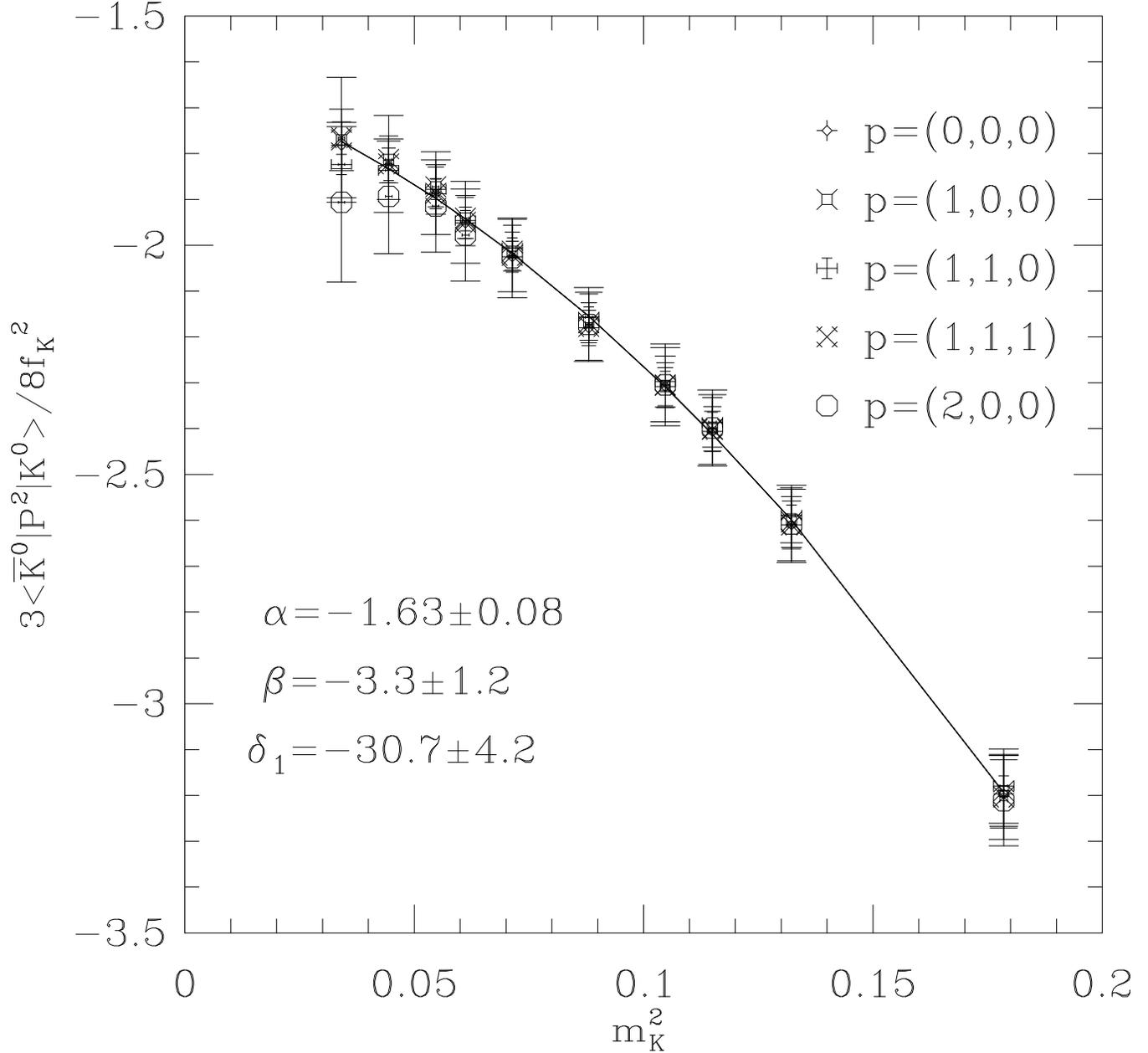}}
%\vskip -24pt plus 10pt
\label{f_P2}
\end{figure}

We close this section with some comments on the reliability of our result.
\begin{enumerate}
\item
Our result is almost independent of the horizontal matching scale $q^*$:
it increases only by 0.02 if we use $q^*=\pi/a$.
This is in apparent contradiction with the conclusions of Ref.~\cite{Bk95APE},
who find a large dependence of $B_K$ on their choice of $\alpha_s$.
The difference is due to the fact that we have used tadpole improved
perturbation theory, in which the perturbative corrections to both
numerator and denominator of $B_K$ are small.
Although the tadpole improvements formally cancel at one-loop order,
there is a significant numerical difference due to  $O(\alpha_s^2)$ terms.
\item
The changes to the coefficients of the subtraction operators required
to cancel the $\alpha$ term in the chiral expansion are smaller than
the one-loop perturbative results. For example, we can cancel $\alpha$ by
increasing the coefficient of $\CO^+_2+\CO^+_3$ from
$-1.0 \alpha_s/\pi$ (its one-loop perturbative value) to 
$-0.6 \alpha_s/\pi \approx - \alpha_s/\pi + 7 (\alpha_s/\pi)^2$.
This is not an unreasonable value for a higher order perturbative term.
\item
Ideally, one would like to determine the matching coefficients
non-perturbatively. First results with Clover action 
\cite{Bk95APE,talevi},
and with Wilson fermions \cite{kuramashi} have been obtained. The 
latter are consistent with our 
result obtained using the chiral fits and perturbative matching. 
\item
Finally, we note that the matrix elements of $\CP^2$ and $\CP^1$,
although they are the largest of all Lorentz structures (see Table
\ref{t_bku2u3}), make only a small contribution to $B_K$.  This is
because they depend, to good approximation, only on $m_K^2$ and not on
$p_i \cdot p_f$ as exemplified in Fig.~\ref{f_P2}.  On the other hand, 
for $B_7^{3/2}$ and $B_8^{3/2}$, where $\CP^2$ and $\CP^1$ are part of
the continuum operator, these operators give the dominant contribution. 
\end{enumerate}

%%% Figure from FIT_NUMBER 6 in  mefitplots.ax (replace data and coefficients)
%%% 4/09/96

\newsection{$B_D$}
\label{sec:BDresults}
 
The analysis of our results for 
\begin{equation}
B_D = { \vev{\bar{D_0}|  \bar c_a \gamma_\mu L u_a \ 
	    \bar c_b \gamma_\mu L u_b | {D_0}} \over
	(8/3)\, \vev{\bar{D_0}| \bar c_a \gamma_\nu \gamma_5 u_a|0}\,
 	      \vev{0| \bar c_b \gamma_\nu \gamma_5 u_b|{D_0}} } 
\,,
\label{eq:bddef}
\end{equation}
and $B_{D_s}$ (related by $u\to s$)
is more straightforward. 
There are no significant constraints from chiral
symmetry---all terms in the expansion of Eq.~(\ref{eq:Bkcpt})
are expected to be present in the physical matrix element.
Thus we simply calculate the $B$-parameters and extrapolate
or interpolate to the physical light quark masses.

Our results for $CS$ and $C U_i$ mesons are given in 
Table~\ref{t_bdmom}. 
They show no significant variation with momentum transfer,
and very little dependence on the light quark mass.
Thus the $\alpha$ term in Eq.~(\ref{eq:Bkcpt}) is dominant---%
in striking contrast to the results for light-light mesons.
Quenched chiral perturbation theory predicts that $B_D$ should diverge
as the light quark mass vanishes \cite{Booth,Zhang}, with the
effect beginning at $m_{s,\rm phys}/3-m_{s,\rm phys}/4$.
We see no sign of such behavior with our light quark masses,
which extend down to approximately $m_{s,\rm phys}/3$.

To investigate the convergence of perturbation theory, we show, in Table~\ref{t_bdmom}, 
separately the ``diagonal'' and ``mixing'' contributions to $B_D$.
The former is the contribution from the lattice operator having the
same form as the continuum operator (i.e. $2 \CO^+_1$), while the
latter is the contribution from the other operators introduced by
one-loop mixing ($\CO^+_{2-5}$).  The mixing contributions are $\sim
15\%$ of $B_D$, indicating reasonable convergence.  Their fractional
contribution is also almost independent of the momentum and the mass
of the light quark.

Our final results are
\begin{equation}
B_D(\mu=q^*=1/a)=0.777(15) \ \ {\rm and}\ \ 
B_{D_s}(\mu=q^*=1/a) = 0.815(9) \,.
\end{equation}
These convert to 
\begin{equation}
B_D(\mu=2\GeV,q^*=1/a)=0.785(15) \ \ {\rm and}\ \ 
B_{D_s}(\mu=2\GeV,q^*=1/a) = 0.822(10) \,.
\end{equation}
The dependence on $q^*$ is, as for $B_K$, smaller than the
statistical errors.
Since $B_D$ and $B_{D_s}$ are correlated, 
their ratio is well determined
\begin{equation}
B_{D_s}/B_D = 1.048(15) \,.
\end{equation}
The variation of $B_D$ with the light quark mass
is described by $0.777(15) + 0.77(27) m_q a$. Here $m_q a$ is the
quark mass obtained from the Ward identity for the axial current, and
$m_s$ is fixed using $M_\phi$ \cite{HM95LANL}.

\begin{table} %4
$$
\let\ifspace=\iffalse
\def\myskip{\omit&&%
%% \omit&&%
%% \omit&&%
%% \omit&&%
%% \omit&&%
%% \omit&&%
 &\cr}
\vbox{\hbox{\vbox{
\tabskip=0pt\offinterlineskip
\def\tlr{\noalign{\hrule}}
\def\CO{{\cal O}}
\def\xx{{\langle \bar \psi \psi \rangle}}
\halign {\strut#& #\tabskip=8pt&
  \hfil$#$\hfil&#&
  \hfil$#$\hfil&#&
  \hfil$#$\hfil&#&
  \hfil$#$\hfil&#&
  \hfil$#$\hfil&#&
  \hfil$#$\hfil&#\tabskip=4pt\cr\tlr
%\omit&height1.5pt&\multispan{11   }&\cr
%
&& 
&& {\ \vec p = (0,0,0)}        
&& {\ \vec p = (1,0,0)}        
&& {\ \vec p = (1,1,0)}        
&& {\ \vec p = (1,1,1)}        
&& {\ \vec p = (2,0,0)}        
  &\cr
\myskip\tlr
% \omit&height0.5pt&\multispan{11   }&\cr\tlr
%
&&\multispan{11   }\hfil {\bf $CS$ } \hfil&\cr\tlr
\myskip
&&  Full 
&&    0.83(  1)
&&    0.84(  1)
&&    0.84(  1)
&&    0.84(  2)
&&    0.85(  2)
  &\cr\ifspace\myskip&&
&& 
&& 
&& 
&& 
&& 
  &\cr\fi\myskip
\myskip
&&  Diagonal
&&    0.71(  1)
&&    0.71(  1)
&&    0.72(  1)
&&    0.72(  1)
&&    0.72(  2)
  &\cr\ifspace\myskip&&
&& 
&& 
&& 
&& 
&& 
  &\cr\fi\myskip
\myskip
&&  Mixing
&&    0.12(  0)
&&    0.12(  0)
&&    0.12(  0)
&&    0.12(  0)
&&    0.12(  0)
  &\cr\ifspace\myskip&&
&& 
&& 
&& 
&& 
&& 
  &\cr\fi\myskip\tlr
&&\multispan{11   }\hfil {\bf $CU_1$ } \hfil&\cr\tlr
\myskip
&&  Full 
&&    0.81(  1)
&&    0.81(  1)
&&    0.81(  1)
&&    0.82(  2)
&&    0.82(  2)
  &\cr\ifspace\myskip&&
&& 
&& 
&& 
&& 
&& 
  &\cr\fi\myskip
\myskip
&&  Diagonal
&&    0.68(  1)
&&    0.69(  1)
&&    0.69(  1)
&&    0.69(  2)
&&    0.69(  2)
  &\cr\ifspace\myskip&&
&& 
&& 
&& 
&& 
&& 
  &\cr\fi\myskip
\myskip
&&  Mixing
&&    0.13(  0)
&&    0.13(  0)
&&    0.13(  0)
&&    0.12(  0)
&&    0.12(  0)
  &\cr\ifspace\myskip&&
&& 
&& 
&& 
&& 
&& 
  &\cr\fi\myskip\tlr
%%%%%%%%%%%%%%
&&\multispan{11   }\hfil {\bf $CU_2$ } \hfil&\cr\tlr
\myskip
&&  Full
&&    0.80(  1)
&&    0.80(  1)
&&    0.80(  2)
&&    0.80(  2)
&&    0.80(  3)
  &\cr\ifspace\myskip&&
&& 
&& 
&& 
&& 
&& 
  &\cr\fi\myskip
\myskip
&&  Diagonal   
&&    0.67(  1)
&&    0.67(  1)
&&    0.68(  1)
&&    0.68(  2)
&&    0.68(  2)
  &\cr\ifspace\myskip&&
&& 
&& 
&& 
&& 
&& 
  &\cr\fi\myskip
\myskip
&&  Mixing   
&&    0.13(  0)
&&    0.13(  0)
&&    0.13(  0)
&&    0.12(  0)
&&    0.12(  0)
  &\cr\ifspace\myskip&&
&& 
&& 
&& 
&& 
&& 
  &\cr\fi\myskip\tlr
%
%%%%%%%%%%%%%%%
&&\multispan{11   }\hfil {\bf $CU_3$ } \hfil&\cr\tlr
\myskip
&&  Full
&&    0.79(  1)
&&    0.79(  2)
&&    0.80(  2)
&&    0.79(  3)
&&    0.79(  3)
  &\cr\ifspace\myskip&&
&& 
&& 
&& 
&& 
&& 
  &\cr\fi\myskip
\myskip
&&  Diagonal
&&    0.67(  1)
&&    0.67(  2)
&&    0.67(  2)
&&    0.67(  3)
&&    0.67(  3)
  &\cr\ifspace\myskip&&
&& 
&& 
&& 
&& 
&& 
  &\cr\fi\myskip
\myskip
&&  Mixing
&&    0.13(  0)
&&    0.13(  0)
&&    0.13(  0)
&&    0.12(  1)
&&    0.12(  1)
  &\cr\ifspace\myskip&&
&& 
&& 
&& 
&& 
&& 1
  &\cr\fi\myskip\tlr
\cr}}}}$$

\caption{Results for $B_D$ (in the TAD1 scheme)
for the heavy-light mesons $CS$, 
$CU_i$ as a function of momentum insertion. 
Also shown are the separate contributions of the diagonal and mixing
parts of the operator.  An error value of $0$ means that it is 
less than $1/2$ in the last significant digit.}
\label{t_bdmom}
\end{table}
%% data from Table on page 19-21 in Bk.tex
%% 4/10/96

\newsection{Results for $B_7^{3/2}$ and $B_8^{3/2}$}
\label{sec:B78results}

We now turn to the left-right operators which appear in the effective
weak Hamiltonian due to electromagnetic penguin diagrams:
\begin{eqnarray}
  \CQ_7 &=&   (\bar s_a \gamma_\mu L d_a) \  \big[ 
                 (\bar u_b \gamma_\mu R u_b) - \half
                 (\bar d_b \gamma_\mu R d_b) - \half
                 (\bar s_b \gamma_\mu R s_b) \big] \,,
\label{eq:Q7}\\
  \CQ_8 &=&      (\bar s_a \gamma_\mu L d_b) \  \big[ 
                 (\bar u_b \gamma_\mu R u_a) - \half
                 (\bar d_b \gamma_\mu R d_a) - \half
                 (\bar s_b \gamma_\mu R s_a) \big] \,.
\label{eq:Q8}
\end{eqnarray}
We have results only for the $I=3/2$ parts of these operators, 
since the $I=1/2$ parts have penguin contractions which we have
not calculated.
Fortunately, the $I=3/2$ parts are of phenomenological interest, 
because they are the only operators giving rise to an imaginary part in
the $K^+\to\pi^+\pi^0$ amplitude, and thus give the dominant electromagnetic
contribution to $\epsilon'/\epsilon$.
We calculate $B$-parameters for $K^+\to\pi^+$ matrix elements,
which are simpler to calculate than those for $K\to\pi\pi$,
but are equal at leading order in chiral perturbation theory.
Thus our $B$-parameters are defined in the continuum by
\begin{eqnarray}
B_7^{3/2} &=& {\vev{\pi^+| \CQ_7^{3/2} | K^+} \over
 \nfrac23 \vev{\pi^+| \bar u \gamma_5 d|0}\vev{0|\bar s\gamma_5 u|K^+}
    - \vev{\pi^+| \bar u \gamma_mu \gamma_5 d|0}\vev{0|\bar s \gamma_\mu \gamma_5 u|K^+} } \,,
\label{eq:b7def}\\
B_8^{3/2} &=& {\vev{\pi^+| \CQ_8^{3/2} | K^+} \over
 2 \vev{\pi^+| \bar u \gamma_5 d|0}\vev{0|\bar s\gamma_5 u|K^+}
 - \nfrac13 \vev{\pi^+| \bar u \gamma_\mu \gamma_5 d|0}\vev{0|\bar s \gamma_\mu \gamma_5 u|K^+} }
\,.
\label{eq:b8def}
\end{eqnarray}
The operators $\CQ_{7,8}^{3/2}$ are defined in Eqs.~(\ref{eq:q7def})
and (\ref{eq:q8def}).

The operators in $B_{7,8}^{3/2}$ are matched onto lattice operators
using the results of Appendix \ref{app:pert}.  A slight complication
arises because the denominators consist of two terms, one of which
(the pseudoscalar) depends on the renormalization scale $\mu$, while
the other (axial) part does not.  Thus we cannot simply use the
$B$-parameters defined above which have only the axial-currents in the
denominator.  Instead, we use horizontal matching followed by 2-loop
running to calculate both numerator and denominator at the final scale
$\mu$, and then take the ratio, all within the jackknife loop.  We
repeat the entire procedure for various choices of $q^*$ and $\mu$.

\begin{figure} 
\hbox{\epsfxsize=\hsize\epsfbox{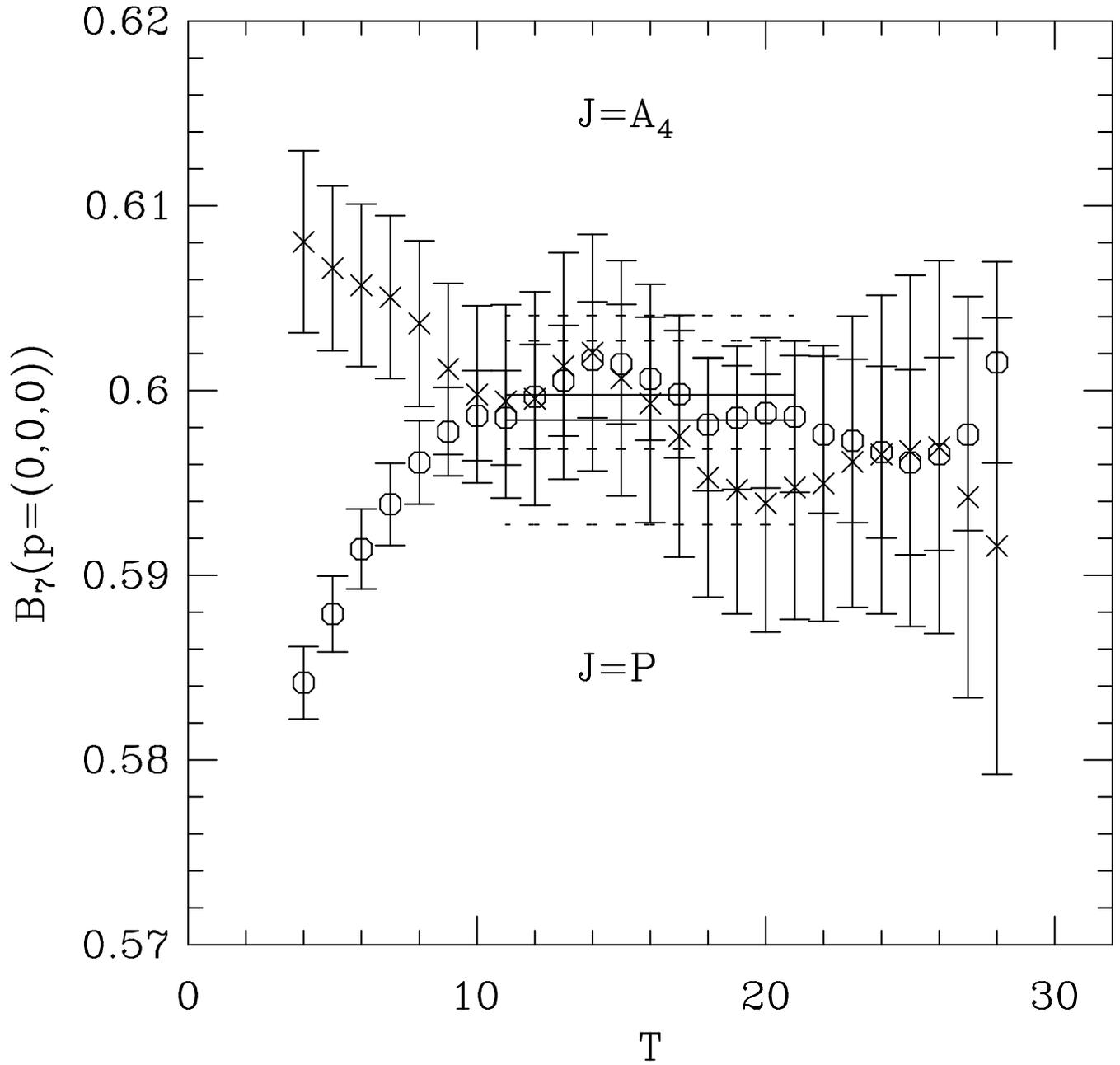}}
\figcaption{Results for the ratio of correlators defining $B_7^{3/2}$,
in TAD1 scheme for the $U_2 U_3$ meson.}
%\vskip -24pt plus 10pt
\label{f_o7fit}
\end{figure}

\begin{figure} %10
\hbox{\epsfxsize=\hsize\epsfbox{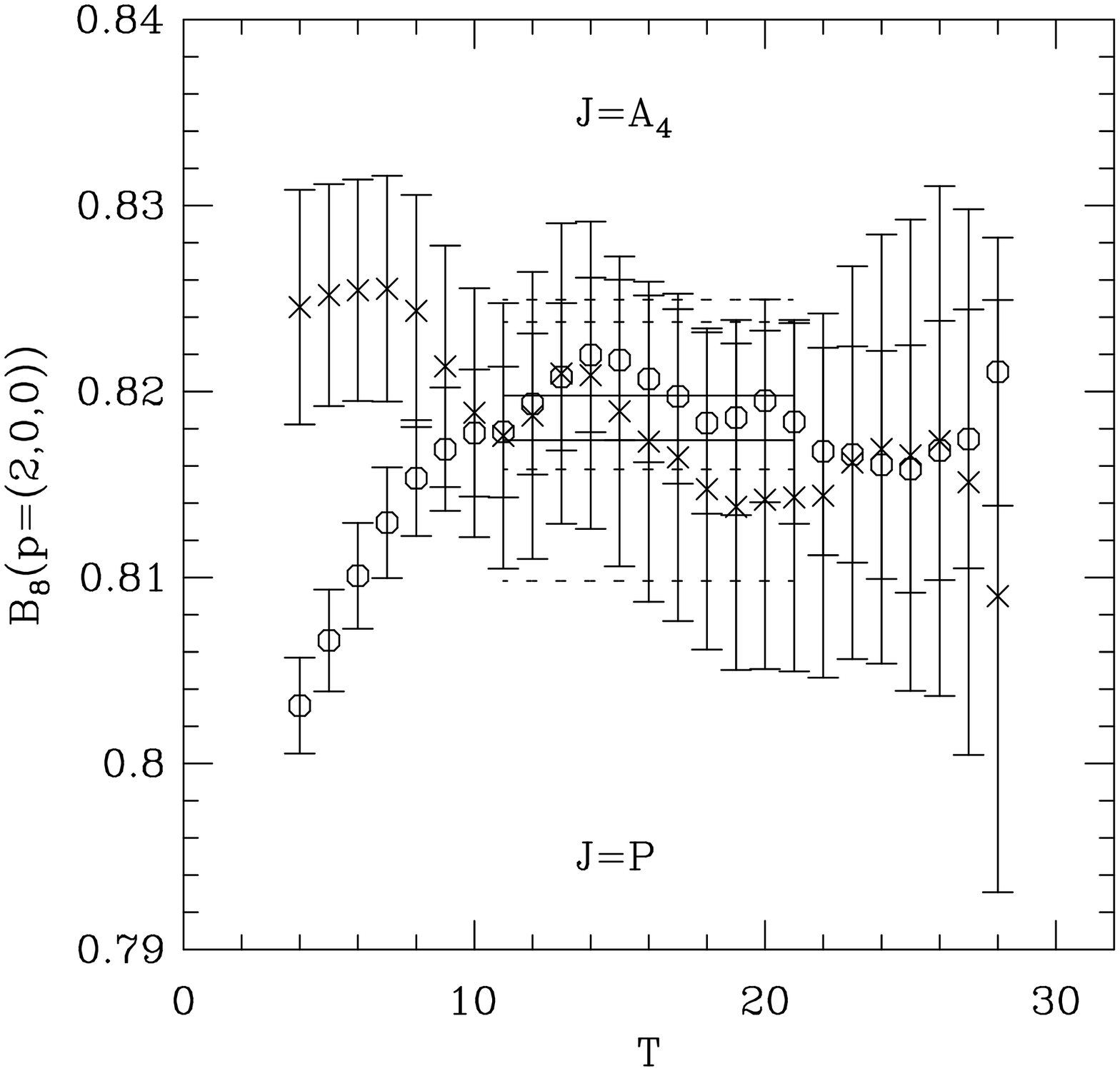}}
\figcaption{Results for $B_8^{3/2}$ in TAD1 scheme for the $U_2 U_3$ meson.}
%\vskip -24pt plus 10pt
\label{f_o8fit}
\end{figure}

The calculation involves no chiral subtractions, since all terms in
the chiral expansion are physical. To show the quality of the results
we display $B_7^{3/2}$ and $B_8^{3/2}$ (in the TAD1 scheme), 
in Figs.~\ref{f_o7fit} and \ref{f_o8fit}, respectively,
for the $U_2 U_3$ meson.
The statistical errors are small, 
and the results using the two sources agree.

%% Generated by outputting from fitting program in.bk.new
%% 4/11/96

%% Generated by outputting from fitting program in.bk.new
%% 4/11/96

We have done the calculation of these matrix elements only for the
case in which the initial and final mesons have the same quark
composition. Our results in the TAD1 scheme are collected in
Table~\ref{t_o7o8mq}.  We see the same gradual increase towards unity
as the meson mass increases that we observed between $B_K$ and $B_D$.
Diagonal and mixing contributions are defined as for $B_D$.
We have also included
results at different momenta for $U_2 U_3$ mesons in the last two rows
of Table~\ref{t_bku2u3}.  The lack of momentum dependence is due to
the dominance of the pseudoscalar matrix elements, which, as noted
above, depend only on $m_K^2$.

\begin{table} %5
\begin{center}
\setlength{\tabcolsep}{8.0pt}
\newcommand\0{\hphantom{0}}
\newcommand\ce[1]{\multicolumn{#1}{c|}}
\newcommand\cf[1]{\multicolumn{#1}{c}}
\begin{tabular}{l|ccc|ccc}
\hline
&\ce3{$B_7$}&\cf3{$B_8$}\cr 
&Full&Diagonal&mixing&Full&Diagonal&mixing\cr 
\hline
$CC    $&$ 0.859(09)$&$ 1.07(1)$&$ -0.22(0)$&$ 0.941(10)$&$ 1.23(1)$&$ -0.30(0)$\cr
$CS    $&$ 0.822(09)$&$ 1.08(1)$&$ -0.26(0)$&$ 0.943(10)$&$ 1.22(2)$&$ -0.29(0)$\cr
$CU_1  $&$ 0.812(10)$&$ 1.07(1)$&$ -0.27(0)$&$ 0.940(11)$&$ 1.20(2)$&$ -0.28(0)$\cr
$CU_2  $&$ 0.806(11)$&$ 1.06(2)$&$ -0.27(0)$&$ 0.936(12)$&$ 1.20(2)$&$ -0.28(1)$\cr
$CU_3  $&$ 0.800(13)$&$ 1.06(2)$&$ -0.27(1)$&$ 0.932(14)$&$ 1.19(3)$&$ -0.28(1)$\cr
$SS    $&$ 0.723(05)$&$ 1.05(1)$&$ -0.34(0)$&$ 0.917(06)$&$ 1.17(1)$&$ -0.26(0)$\cr
$SU_1  $&$ 0.695(04)$&$ 1.02(1)$&$ -0.34(0)$&$ 0.900(05)$&$ 1.14(2)$&$ -0.25(0)$\cr
$SU_2  $&$ 0.680(04)$&$ 1.01(2)$&$ -0.35(1)$&$ 0.890(06)$&$ 1.12(2)$&$ -0.25(0)$\cr
$SU_3  $&$ 0.669(05)$&$ 1.00(2)$&$ -0.34(1)$&$ 0.880(06)$&$ 1.11(2)$&$ -0.25(1)$\cr
$U_1U_1$&$ 0.660(04)$&$ 0.99(2)$&$ -0.35(1)$&$ 0.874(05)$&$ 1.10(2)$&$ -0.24(1)$\cr
$U_1U_2$&$ 0.640(04)$&$ 0.97(2)$&$ -0.35(1)$&$ 0.858(05)$&$ 1.08(3)$&$ -0.24(1)$\cr
$U_1U_3$&$ 0.625(04)$&$ 0.96(3)$&$ -0.34(1)$&$ 0.843(05)$&$ 1.06(3)$&$ -0.23(1)$\cr
$U_2U_2$&$ 0.617(04)$&$ 0.95(3)$&$ -0.34(1)$&$ 0.837(05)$&$ 1.05(3)$&$ -0.23(1)$\cr
$U_2U_3$&$ 0.599(04)$&$ 0.93(3)$&$ -0.34(1)$&$ 0.819(06)$&$ 1.03(3)$&$ -0.23(1)$\cr
$U_3U_3$&$ 0.578(05)$&$ 0.91(4)$&$ -0.34(1)$&$ 0.797(07)$&$ 1.01(4)$&$ -0.22(1)$\cr
\hline
\end{tabular}

\end{center}
\caption{Results for $B_7^{3/2}$ and $B_8^{3/2}$ in the TAD1 scheme 
($\mu=q^*=1/a$).}
%\vskip 6pt
\label{t_o7o8mq}
\end{table}

In Table~\ref{t_o7o8final} we give results for $\mu=2 \GeV$ with
various values of $q^*$. There is a significant variation with $q^*$,
considerably larger than the statistical errors.
This indicates that the perturbative expansion for the
anomalous dimensions is converging more slowly than for $B_K$ and $B_D$.
We take $q^*=1/a$ for our final result, and use the range of values for
$q^*=2 \GeV-\pi/a$ to estimate the systematic error. This leads to
\begin{eqnarray}
  B_7^{3/2}(NDR,2\GeV) &=&  0.58 \pm 0.02 ({\rm stat})
				{+0.07 \atop -0.03} ({\rm pert}) \,,  \\
  B_8^{3/2}(NDR,2\GeV) &=&  0.81 \pm 0.03 ({\rm stat}) 
				{+0.03 \atop -0.02} ({\rm pert}) \,.
\label{eq:o7o8numbers}
\end{eqnarray}
%%% NEW---OK, LETS KEEP THIS. STEVE
Note that the statistical errors are somewhat larger after running to
$\mu=2\GeV$ (compare results in Table~\ref{t_o7o8final} to those in
Table~\ref{t_o7o8mq}).  This is because, in order to evaluate the
$B-$parameters at $2 \GeV$, one has to combine results of three
separate fits after the running. Consequently, the cancellation of errors
between the numerator and denominator is not as good.

\begin{table} %6
\begin{center}
\begin{tabular}{lcc}
\hline
$q^*$ 		&$B_7^{3/2}$	& $B_8^{3/2}$ \\
\hline
$2 \GeV$	&$ 0.549(18)$	&$ 0.790(27)$\\
$1/a$       	&$ 0.578(19)$	&$ 0.807(27)$\\
$2/a$       	&$ 0.641(21)$	&$ 0.835(28)$\\
$\pi/a$	     	&$ 0.654(21)$	&$ 0.837(28)$\\
\hline
\end{tabular}
\caption{Variation of $B_7$ and $B_8$, run to $\mu=2 \GeV$, 
for various choices of $q^*$.}
\label{t_o7o8final}
\end{center}
\end{table}
%% data from Table on page 14 in auxtabs.tex
%% 4/11/96

\newsection{Results for $B_S$}
\label{sec:BSresults}

We close with results for the $B$-parameter introduced in 
Ref.~\cite{beneke} to calculate the difference of B-meson lifetimes
\begin{equation}
B_S \equiv B_4^+= {\vev{B_s|\bar b_a L s_a \, \bar b_c L s_c| \bar{B_s}}  \over
(5/3) \vev{B_s|\bar b_a \gamma_5 s_a|0} \vev{0|\bar b_a \gamma_5 s_a|\bar{B_s}
}}
\,.
\label{eq:bsdef}
\end{equation}
We also consider the operator related by a Fierz transformation
\begin{equation}
B_5^+ = {\vev{B_s|\bar b_a L s_c \, \bar b_c L s_a| \bar{B_s}}  \over
(-1/3) \vev{B_s|\bar b_a \gamma_5 s_a|0} \vev{0|\bar b_a \gamma_5 s_a|\bar{B_s}
}}
\,.
\label{eq:b5def}
\end{equation}
The perturbative matching coefficients for this operator have not been
calculated previously---they are given in Appendix \ref{app:pert}.
The 1-loop improved operators $O_4^+$ and $O_5^+$ are defined in
Eqs.~\ref{e:O45defs} and \ref{e:VSA45defs}. (One needs the matrix
elements of both operators due to the mixing between them.)  The
calculation of their matrix elements and the corresponding
B-parameters is straightforward, and the results for the 15 mass
combinations, as a function of the matching scale $\mu=q^*$, are given
in Tables~\ref{t_O4mq} and \ref{t_O5mq}.
The results after extrapolation of the light quark to the physical
strange quark mass are given in Table~\ref{t_O45qstar}.

Note that, unlike the previous cases of $B$-parameters, we cannot run
these results to a final common scale.  As can be seen from the
results of Appendix A, this requires knowledge of the two-loop
anomalous dimensions.  It would be inconsistent to use one-loop
matching and then only run using one-loop anomalous dimensions.  Thus
we cannot study the $q^*$ dependence of the results at fixed final
scale $\mu$.  The fact that both $B_{4^+}$ and $B_{5^+}$, evaluated at
$\mu=q^*$, depend only weakly on $q^*$, as is apparent from the
Tables, is an accident.  The origin of this feature can be traced to
the large value of the matrix elements of the lattice mixing
operators, and consequently significant variation with $\alpha_s(q^*)$.

The best we can do at present is to use our preferred value of $q^*=1/a$,
and quote the results only for $\mu=q^*$, i.e. using horizontal matching.
These results are
\begin{eqnarray}
  B_4^{+}(\NDR,\mu=q^*=1/a) &=&  0.80 \pm 0.01 ({\rm stat})  \,,  \\
  B_5^{+}(\NDR,\mu=q^*=1/a) &=&  0.94 \pm 0.01 ({\rm stat})  \,.
\label{eq:o4o5numbers}
\end{eqnarray}

\begin{table} %7
\begin{center}
\setlength{\tabcolsep}{3.0pt}
\newcommand\1{\hphantom{1}}
\begin{tabular}{l|cccc}
\hline
 &$q^*=2 \GeV$
 &$q^*=1/a$
 &$q^*=2/a$
 &$q^*=\pi/a$
 \cr 
\hline
$CC    $&$  0.877(9)$&$  0.883(9)$&$  0.880(9)$&$  0.871(9)$ \cr
$CS    $&$  0.806(6)$&$  0.813(6)$&$  0.813(6)$&$  0.806(6)$ \cr
$CU_1  $&$  0.784(8)$&$  0.791(8)$&$  0.792(8)$&$  0.786(8)$ \cr
$CU_2  $&$  0.771(8)$&$  0.779(8)$&$  0.781(8)$&$  0.775(8)$ \cr
$CU_3  $&$  0.763(9)$&$  0.771(9)$&$  0.773(9)$&$  0.767(9)$ \cr
$SS    $&$  0.696(4)$&$  0.703(4)$&$  0.707(4)$&$  0.701(4)$ \cr
$SU_1  $&$  0.663(3)$&$  0.670(4)$&$  0.675(4)$&$  0.670(4)$ \cr
$SU_2  $&$  0.647(4)$&$  0.655(4)$&$  0.660(4)$&$  0.656(4)$ \cr
$SU_3  $&$  0.639(4)$&$  0.646(4)$&$  0.652(4)$&$  0.647(4)$ \cr
$U_1U_1$&$  0.626(3)$&$  0.634(3)$&$  0.640(3)$&$  0.636(3)$ \cr
$U_1U_2$&$  0.610(3)$&$  0.617(3)$&$  0.623(3)$&$  0.619(3)$ \cr
$U_1U_3$&$  0.600(4)$&$  0.607(4)$&$  0.613(4)$&$  0.609(4)$ \cr
$U_2U_2$&$  0.592(4)$&$  0.599(4)$&$  0.605(4)$&$  0.601(4)$ \cr
$U_2U_3$&$  0.581(5)$&$  0.588(4)$&$  0.594(4)$&$  0.590(4)$ \cr
$U_3U_3$&$  0.571(6)$&$  0.578(6)$&$  0.583(6)$&$  0.579(5)$ \cr
\hline
\end{tabular}

\caption{The B-parameter for the operator ${O_4^+}$. 
The data are shown for all 15 mass combinations, and for various choices 
of $\mu=q^*$.}
\label{t_O4mq}
\end{center}
\end{table}
%% extracted by hand from auxtabs.tex 

\begin{table} % Table 8
\begin{center}
\setlength{\tabcolsep}{3.0pt}
\newcommand\1{\hphantom{1}}
\begin{tabular}{l|cccc}
\hline
 &$q^*=2 \GeV$
 &$q^*=1/a$
 &$q^*=2/a$
 &$q^*=\pi/a$
 \cr 
\hline
$CC    $&$  0.975(  11)$&$  0.991(  11)$&$  1.013(  11)$&$  1.011(  11)$ \cr
$CS    $&$  0.940(\1 9)$&$  0.953(\1 9)$&$  0.967(\1 9)$&$  0.963(\1 9)$ \cr
$CU_1  $&$  0.920(  10)$&$  0.933(  10)$&$  0.946(  10)$&$  0.942(  10)$ \cr
$CU_2  $&$  0.910(  12)$&$  0.922(  11)$&$  0.935(  11)$&$  0.931(  11)$ \cr
$CU_3  $&$  0.902(  14)$&$  0.914(  13)$&$  0.927(  12)$&$  0.923(  12)$ \cr
$SS    $&$  0.878(\1 5)$&$  0.885(\1 5)$&$  0.885(\1 5)$&$  0.878(\1 5)$ \cr
$SU_1  $&$  0.850(\1 5)$&$  0.856(\1 5)$&$  0.855(\1 5)$&$  0.847(\1 5)$ \cr
$SU_2  $&$  0.836(\1 5)$&$  0.842(\1 5)$&$  0.840(\1 5)$&$  0.832(\1 5)$ \cr
$SU_3  $&$  0.829(\1 6)$&$  0.835(\1 6)$&$  0.832(\1 5)$&$  0.824(\1 5)$ \cr
$U_1U_1$&$  0.817(\1 4)$&$  0.822(\1 4)$&$  0.819(\1 4)$&$  0.811(\1 4)$ \cr
$U_1U_2$&$  0.802(\1 5)$&$  0.806(\1 5)$&$  0.802(\1 4)$&$  0.794(\1 4)$ \cr
$U_1U_3$&$  0.793(\1 5)$&$  0.797(\1 5)$&$  0.792(\1 5)$&$  0.783(\1 5)$ \cr
$U_2U_2$&$  0.785(\1 5)$&$  0.788(\1 5)$&$  0.783(\1 5)$&$  0.774(\1 5)$ \cr
$U_2U_3$&$  0.775(\1 6)$&$  0.778(\1 6)$&$  0.771(\1 6)$&$  0.763(\1 6)$ \cr
$U_3U_3$&$  0.765(\1 8)$&$  0.768(\1 8)$&$  0.759(\1 7)$&$  0.750(\1 7)$ \cr
\hline
\end{tabular}

\caption{The B-parameter for the operator ${O_5^+}$. The data are
shown for all 15 mass combinations, and for various choices of
$\mu=q^*$.}
\label{t_O5mq}
\end{center}
\end{table}
%% extracted by hand from auxtabs.tex 

\begin{table} % Table 9
\begin{center}
\setlength{\tabcolsep}{6.0pt}
\newcommand\ce[1]{\multicolumn{#1}{c}}
\begin{tabular}{l|c|c}
\hline
 &{$B_4^+(\mu=q^*)$}
 &{$B_5^+(\mu=q^*)$}
 \cr 
 \hline
$q^*=2 \GeV$     &$0.791(8)$&$0.927(10)$ \cr
$q^*=1/a  $      &$0.798(8)$&$0.940(10)$ \cr
$q^*=2/a  $      &$0.799(8)$&$0.953(10)$ \cr
$q^*=\pi/a$      &$0.792(8)$&$0.949(10)$ \cr
\hline
\end{tabular}

%% \setlength{\tabcolsep}{6.0pt}
%% \newcommand\ce[1]{\multicolumn{#1}{c}}
%% \begin{tabular}{l|cc|cc}
%% \hline
%% \ce1{ }&\ce2{$B_{4^+}$}&\ce2{$B_{5^+}$}\cr 
%%   \ce1{ }
%%  &\ce1{$\mu=q^*   $}
%%  &\ce1{$\mu=2 \GeV$}
%%  &\ce1{$\mu=q^*   $}
%%  &\ce1{$\mu=2 \GeV$}
%%  \cr 
%%  \hline
%% $q^*=2 \GeV$     &$0.791(8)$&$0.791(09)$&$0.927(10)$&$0.927(10)$ \cr
%% $q^*=1/a  $      &$0.798(8)$&$0.834(09)$&$0.940(10)$&$0.982(11)$ \cr
%% $q^*=2/a  $      &$0.799(8)$&$0.991(11)$&$0.953(10)$&$1.182(14)$ \cr
%% $q^*=\pi/a$      &$0.792(8)$&$1.077(12)$&$0.949(10)$&$1.290(15)$ \cr
%% \hline
%% \end{tabular}

\caption{The B-parameter for the operator ${O_4^+}$ and ${O_5^+}$
after horizontal matching to the continuum at scale $q^*$. 
%The lattice scale $1/a=2330(41) \MeV$.  
The $CU_i$ data have been
extrapolated to $m_s(M_\phi)$ in the light quark mass.}
\label{t_O45qstar}
\end{center}
\end{table}
%% extracted by hand from auxtabs.tex 

\section*{Acknowledgements}

These calculations have been done on the CM5 at LANL as part of the
DOE HPCC Grand Challenge program, and at NCSA under a Metacenter
allocation.  We thank Jeff Mandula, Larry Smarr, Andy White and the
entire staff at the two centers for their tremendous support
throughout this project.

\appendix

\newsection{Issues in matching lattice and continuum operators}
\label{app:match}

In this appendix we discuss the matching of
lattice and continuum operators.
We also collect results for the anomalous dimensions needed 
for the particular matrix elements we calculate.

We wish to find the lattice regularized operators which match with
(i.e. have the same physical matrix elements as) the continuum
operators of interest:
\begin{equation}
\CO^{\rm cont}(\mu) = Z(\mu,a) \CO^{\rm latt}(a) \,.
\end{equation}
Here $\CO^{\rm cont}$ is a vector of operators which is closed
under continuum and lattice mixing, and $\CO^{\rm latt}$ the corresponding
vector of lattice operators.
By comparing matrix elements, one can calculate the matrix 
$Z$ in perturbation theory.
The one-loop result has the form
\begin{equation}
Z(\mu, a) = 1 + \lambda\, c(\mu a) \,,
\label{eq:naivematch}
\end{equation}
where $\lambda=g^2/(16 \pi^2)$ and $c$ is the one-loop matching matrix.
Appendix \ref{app:pert} describes the calculation of $c$ for four-fermion
operators. The issue we address here is the
choice of $g$ to use in $\lambda$.
Should it be the continuum coupling $g_{\overline{\rm MS}}(\mu)$,
or the bare lattice coupling $g(a)$, or some sort of an average?
The uncertainty this introduces can be comparable to the statistical error,
as is the case in the calculation of $B_K$ using staggered 
fermions~\cite{sharpebk94}.

To address this issue we use the exact
perturbative formula for $Z$ given by Ji~\cite{Ji}.
Generalized to the case of mixing, this is
\begin{equation}
Z(\mu,a) =
T_{g''} \exp\left(- \int_0^{g(\mu)} dg''
{\gamma_{\rm cont}(g'') \over \beta_{\rm cont}(g'')}\right)
\ T_{g'} \exp\left(- \int_{g(a)}^0 dg' 
       {\gamma_{\rm latt}(g') \over \beta_{\rm latt}(g')} \right)
\,.
\label{eq:Jimatch1}
\end{equation}
The $\gamma$'s are the anomalous dimension matrices, which are different
in the continuum and on the lattice. 
The $\beta$'s are the usual $\beta$-functions, again different
in the continuum and on the lattice, though only at $O(g^7)$.
The $T_g$'s indicate that the exponential integrals are $g$-ordered.
This formula has a simple physical interpretation:
first use the renormalization group to
run on the lattice to a lattice spacing as small as one wants,
then match with the continuum at tree level,
and finally run back to the desired scale in the continuum theory.
It is exact except for possible non-perturbative corrections.

To use Ji's formula we have to specify both the continuum and lattice
renormalization schemes. In the continuum we take NDR, 
(i.e. $\MSbar$ plus a particular set of rules for dealing with
$\gamma_5$ in $n$ dimensions)
so that $g(\mu)$ is in the $\MSbar$ scheme.
On the lattice, Ji uses the bare lattice coupling constant, 
and takes the scale to be $a$.
The work of Lepage and Mackenzie suggests, however, that it is better to
reexpress all lattice perturbative expressions in terms of a 
continuum-like coupling \cite{Lepage}.
Thus we suggest improving Ji's formula by using the same definition
of renormalized coupling constant on the lattice as in the continuum.
%Of course, the regularization of the operators is still done by the lattice. 
In other words, we imagine expressing perturbative series on the lattice,
e.g. that for the anomalous dimension matrix,
in terms of $g_{\overline{\rm MS}}(q^*)$,
where $q^* a = K$, with $K$ a constant.
This is just a change of variables
\begin{equation}
g_{\overline{\rm MS}}(q^*,g(a)) = g(a) + c_1(q^* a) g(a)^3 + O(g^5)
\,,
\end{equation}
%%% NEW
with $c_1$ a known constant with known dependence on $q^* a$.
The value of $q^* a$ should be chosen so as to improve the
convergence of perturbative series.

To implement this improvement, we change variables
in the integral over $g'$ in Eq.~(\ref{eq:Jimatch1}),
from $g'$ to $g_{\overline{\rm MS}}(q^*=K/a,g')$.
The result is
\begin{eqnarray}
\int_{g(a)}^0 dg' {\gamma_{\rm latt}(g') \over \beta_{\rm latt}(g')} 
&=&  \int^0_{g_{\overline{\rm MS}}(q^*=K/a)}
{dg_{\overline{\rm MS}} \over |dg_{\overline{\rm MS}}(g')/dg'|} \ 
{\gamma_{\rm latt}(g'(g_{\overline{\rm MS}})) \over (-dg'(a)/d\ln a)} \\
&=&  \int^0_{g_{\overline{\rm MS}}(q^*=K/a)} 
dg_{\overline{\rm MS}} \ 
{\gamma_{\rm latt}(g'(g_{\overline{\rm MS}})) \over 
dg_{\overline{\rm MS}}/d\ln q^*} \\
&=&  \int^0_{g_{\overline{\rm MS}}(q^*=K/a)}
dg_{\overline{\rm MS}} \ 
{\gamma_{\rm latt}(g'(g_{\overline{\rm MS}})) \over 
\beta_{cont}(g_{\overline{\rm MS}})} \,.
\label{eq:newintegral}
\end{eqnarray}
Thus we end up with the same form, but with the continuum $\beta$-function
%%% NEW
in the denominator, a different lower limit, and the anomalous dimension
matrix expressed in terms of $g_{\overline{\rm MS}}$.  This change of
variables makes no difference if we work to all orders, but does
affect the result when we truncate perturbation theory.  The hope is
that $\gamma_{\rm latt}$ converges more quickly as a result of this
reorganization.

In practice we know only the two-loop anomalous dimensions for the
operators of interest, so we expand out Eq.~\ref{eq:Jimatch1}
%%% NEW
(with the substitution \ref{eq:newintegral}) to second
order, following the method of Ref.~\cite{burasREV}
\begin{equation}
Z(\mu,a) = 
[1 + \lambda(\mu) J_{\rm cont} + O(\lambda^2)]^{-1}\,  W^{-1} \,
	\left(\lambda(\mu) \over \lambda(q^*)\right)^{\gamma_0^D / 2\beta_0}
	\, W\, [1 + \lambda(q^*) J_{\rm latt}(q^*) + O(\lambda^2)]
\,.
\label{eq:Jimatch2}
\end{equation}
Note that the $g^2$ in $\lambda$ is in the $\MSbar$ scheme 
wherever it appears. Note also that the $a$ dependence comes in implicitly
through $q^*$.
$\gamma^D_0$ is the result of diagonalizing the
the one-loop anomalous dimension matrix, $\gamma_0$,
(which is the same in both lattice and continuum schemes)
\begin{equation}
	\gamma^D_0 = W \gamma_0 W^{-1} \,.
\end{equation}
The 2-loop contribution feeds in through
\begin{equation}
 J = {\beta_1 \gamma_0 \over 2\beta_0^2}  - W^{-1} M W \ .
\end{equation}
where the matrix $M$ is 
\begin{equation}
 M_{ij} \ = \ {\big( W \gamma_1 W^{-1} \big)_{ij} \over 
    2\beta_0 - \big( \gamma^D_0 \big)_{ii} + \big( \gamma^D_0 \big)_{jj} } \ .
\end{equation}
$J_{\rm cont}$ differs from $J_{\rm latt}$ because the
two-loop anomalous dimension matrices differ.
In practice, the lattice $J$'s contain more off-diagonal terms than
those in the continuum.
For example, in the continuum the operators
$\CQ_7$ and $\CQ_8$ only mix with each other
(and with evanescent operators which vanish in 4-dimensions),
while $J_{\rm latt}$ mixes these operators also with $\CO^+_{1,4,5}$,
as described in Appendix \ref{app:pert}.
The structure of Eq.~\ref{eq:Jimatch2} means, however, that,
working from left to right, we can stay in the two dimensional
space spanned by $(\CQ_7,\CQ_8)$ until we encounter $J_{\rm latt}$.
Thus we need only the rectangular piece of $J_{\rm latt}$
which connects $(\CQ_7,\CQ_8)$ to the full basis.

Equation (\ref{eq:Jimatch2}) is (aside from a minor modification
discussed below) the result which we propose using in place
of Eq.~(\ref{eq:naivematch}).
It makes clear which scale to evaluate the coupling at wherever it appears:
the lattice scale in the lattice anomalous dimension,
and the continuum scale in the continuum anomalous dimension.
It is valid even if $\mu a$ is substantially different from unity,
for it sums up the large logarithms of $\mu a$.
It can be applied to tadpole improved operators---tadpole improvement
simply changes the two-loop lattice anomalous dimension, 
presumably making the perturbative series converge more quickly.

To apply Eq.~\ref{eq:Jimatch2} we need to know the $J$'s.
In the continuum, the two-loop anomalous dimensions are known 
for most of the operators of interest (they are listed below),
and from these we can construct $J_{\rm cont}$.
$J_{\rm latt}$ can then be determined by comparing the
one-loop matching result to the general formula, in the following way.
We recall that, when deriving Eq.~(\ref{eq:naivematch}), 
one equates the lattice and continuum coupling constants 
(their difference being a higher order effect).
For this to be true in the general formula we must set $\mu=q^*$.
Thus we have
\begin{eqnarray}
Z(\mu=q^*,a) &\equiv& [1 + \lambda(q^*)\;c(q^* a) + O(\lambda^2) ]\nonumber \\
&=&
[1 + \lambda(q^*) J_{\rm cont} + O(\lambda^2)]^{-1}\,
[1 + \lambda(q^*) J_{\rm latt}(q^*) + O(\lambda^2)] \,,
\label{eq:Jimatch3}
\end{eqnarray}
and so
\begin{equation}
J_{\rm latt}(q^*) = J_{\rm cont} + c(q^* a) \,.
\end{equation}
We now see explicitly that $J_{\rm latt}$ depends on $q^*$,
and upon whether we tadpole improve or not (which changes 
$c \to c-{\rm tad}$, where {\rm tad} is the tadpole contribution).

To express the result at scale $\mu$, we have in fact used a slight
variant of Eq.~\ref{eq:Jimatch2}, which we call ``horizontal
matching'':
\begin{equation}
Z(\mu,a) = 
[1 + \lambda(\mu) J_{\rm cont}]^{-1}\,  W^{-1} \,
	\left(\lambda(\mu) \over \lambda(q^*)\right)^{\gamma_0^D\over 2\beta_0}
	\, W\, [1 + \lambda(q^*) J_{\rm cont}]
               [1 + \lambda(q^*) c(q^*a)]
\,.
\label{eq:horizontal}
\end{equation}
This differs by corrections of size $\lambda(q^*)^2$,
which turn out to be much smaller than our statistical errors.
The physical interpretation of horizontal matching is that we first match
lattice and continuum operators at the scale $q^*$
(using the same coupling for both), and then run, in the continuum, down
to the final scale $\mu$.
The important practical advantage of 
both horizontal matching and that based on
Ji's approach is the improved treatment of terms involving $\lambda$,
when $\lambda(\mu)$ is significantly different from $\lambda(q^*)$.
We have, in effect, used the renormalization group to
include a subset of $O(\lambda^2)$ terms.

The remaining issue is the choice of $q^*$.
That we have such a choice simply reflects the fact that we have
truncated the perturbative expansions for $\gamma$ and $\beta$.
Lepage and Mackenzie have suggested a scheme for estimating $q^*$ 
which works for finite lattice renormalizations~\cite{Lepage}.
We have not, however,
found a way to implement their scheme for divergent operators.
Presumably one could use the alternative BLM scheme~\cite{BLM},
but this involves calculating a subset of diagrams for the two-loop
matching coefficients (or three-loop anomalous dimensions), 
which has not been done.
Thus we have simply carried
out the calculation for a reasonable range of $q^*$:
$2 \GeV$, $1/a$, $2/a$ and $\pi/a$.
Previous work suggests that, for tadpole improved operators,
the optimum $q^*$ lies towards the lower end of this range \cite{Lepage}.
For this reason we quote our final value using $q^*=1/a$.
It is important to realize, however, that this intuition concerning
$q^*$ may not apply to matching calculations such as those we are using here.
Thus we use the results from the entire range of values of $q^*$ to
estimate a systematic error.

We close this appendix by collecting results needed 
to carry out the matching just described.
We quote all results for $n_f=0$, and give
anomalous dimensions in the $\NDR$ scheme.
The operators are listed in Eqs.~(\ref{eq:qdef}-\ref{eq:qsdef}).

The coefficients $\beta_0$ and $\beta_1$ of the $\beta-$function 
($\mu \partial g / \partial \mu = -\beta_0 g^3 - \beta_1 g^5 - \ldots$)
are independent of the renormalization scheme, and are standard.
% For the $n_f = 0$ theory 
%\begin{equation}
%\beta_0 =  11\,, \qquad \beta_1 =  102 \,.
%\end{equation}
The anomalous dimensions for $\CQ$ are 
\begin{equation}
\gamma_0 =  4 \,,\qquad \gamma_1 =  -7 \,.
\end{equation}
The operators $\CQ_7^{3/2}$ and $\CQ_8^{3/2}$ mix, with anomalous dimensions
\begin{eqnarray}
\gamma_0 &=&  \left( \begin{array}{cc}
                2\ \  & -6 \\
                0\ \  & -16
                \end{array} \right) \,,\nonumber \\
\gamma_1 &=&  \left( \begin{array}{cc}
                71/3 \ &\ -99 \\
               -225/2\ &\ -1331/6
                \end{array} \right) \,.
\end{eqnarray}
The result for $\gamma_0$ is standard, while we have
extracted $\gamma_1$ from the results of Ref.~\cite{Burasetal}
by considering only a subset of the graphs required 
for the full operators $\CQ_{7,8}$. The results for the operators 
needed to calculate $B_s$ are given at the end of Appendix~\ref{app:pert}. 

% appert.tex --- appendix on perturbation theory
\newsection{One-loop matching for four-fermion operators}
\label{app:pert}

In this appendix we give a general formula for the one loop matching
coefficients for four-fermion operators of the form $\Gamma\otimes\Gamma$.
This formula relates continuum operators defined in the NDR
%(naive dimensional regularization) 
scheme to lattice operators which are local, 
i.e. reside on a single lattice site.
All one needs to know are the matching coefficients for the five
bilinears $S$, $P$, $V$, $A$ and $T$.
The formula applies not only for Wilson fermions, but 
for any improved Wilson fermion action, e.g. the Sheikholeslami-Wohlert action.
It is not useful for staggered fermions, where many of the operators 
of interest are not local (see, for example, Ref.~\cite{SharpePatel}).

Matching coefficients for a subset of four-fermion operators were
calculated long ago by Martinelli \cite{pertmarti} and 
Bernard {\sl et al.} \cite{pertucla}.
These authors use different regularization schemes in the continuum,
namely DRED (dimensional reduction) and DR$\overline{\rm EZ}$ 
(a variant of dimensional reduction).
One of our purposes here is to convert these results to the NDR scheme,
which is the standard scheme used in the evaluation of continuum
coefficient functions.
%Results allowing this conversion have been worked out for staggered
%fermions\cite{SharpePatel}, but since this is a conversion between
%continuum schemes, the results apply just as well to 
%(improved) Wilson fermions.
A second purpose is to extend the calculations to operators
of the form $S\otimes S + P\otimes P$ and $T\otimes T$.
%results for which have not been given previously.
As discussed in the text, the matrix elements of one such operator
has recently been shown to be of phenomenological interest.
%Thus it is desirable to
%complete the calculation of matching coefficients for all
%Lorentz scalar four-fermion operators.
Our final purpose is to present all the results in a simple form
which is easy to evaluate.

Our method is adapted from that used for staggered 
fermions in Ref.~\cite{SharpePatel}.
It proceeds in two stages. 
The first stage uses the method of Martinelli \cite{pertmarti},
who pointed out that all the one-loop vertex diagrams for four-fermion
operators can be brought into the form of bilinear corrections using
Fierz transformations and charge conjugation.
This works, however, only if the continuum scheme keeps the gamma-matrices
in four dimensions, which is why he chose DRED.
Following Ref.~\cite{SharpePatel},
we actually use a slightly different intermediate scheme, ${\rm NDR}'$.
%which is more easily related to NDR.
In the second stage we convert from ${\rm NDR}'$ to NDR.
This part of the calculation is done in the continuum and it cannot
be reduced to that for bilinears.
It is, however, independent of the lattice fermion action.
Much of the work for the second stage has been done in
Ref.~\cite{SharpePatel}.
Our only new result is the construction of an
$n$-dimensional definition of the operators
$S\otimes S +P\otimes P$ and $T\otimes T$ which maintains
Fierz symmetries. This is not an essential condition, but it is 
desirable for aesthetic reasons.

We begin with some notation.
The one-loop matching coefficients for bilinears are defined by
\begin{equation}
\CB^{\rm cont}_i = \CB^{\rm lat}_i 
	\left( 1 + C_F\ \lambda \ c_i(\mu a)\right) \,,
\end{equation}
where $i=S,V,T,A,P$,
$C_F=4/3$, $\lambda=g^2/16 \pi^2$, $\mu$ is the renormalization
scale of the continuum operator
and $a$ the lattice spacing.
%The value of $\lambda$ to be used in this equation is discussed in the
%text---it is about $1/65$ for our simulation.
The values of the $c_i$ are given in Table~\ref{tab:bilincorr} for
Wilson fermions in both DRED and NDR schemes.
These values are taken from Refs. \cite{pertmarti,pertucla},
and converted to NDR using the conversion factors given
in Ref. \cite{PatelSharpe}.
Note that the corrections are uniformly small after tadpole improvement
\cite{Lepage}.  For completeness we mention that 
the analogous results of the renormalization constants for bilinears with 
the Sheikholeslami-Wohlert (clover) action have been calculated 
by G\"ockeler \etal \cite{ZforSW}. 

\begin{table} %4
\renewcommand{\arraystretch}{1.2}
\begin{center}
\begin{tabular}{llccccc}
\hline
Action & Scheme
		& $c_S$	& $c_V$	& $c_T$ & $c_A$	& $c_P$ \\
\hline
Wilson & NDR	&$-12.952 + 6 \ell$	
			&$-20.618$ 
				&$-17.01 - 2 \ell$
					&$-15.796$
						&$-22.596 + 6 \ell$ \\
Wilson & DRED${}-{}$NDR
		& 1	& $0.5$	& -1	& $0.5$	& 1 \\
\hline
\end{tabular}
\caption{Bilinear matching coefficients,
with $\ell = \log(\mu a)$.
Tadpole improvement (discussed in the main text) adds
${\rm tad}/C_F=12.86$ to each of these numbers.}
\label{tab:bilincorr}
\end{center}
\end{table}

The four-fermion operators we consider are linear combinations of
the five Lorentz scalars
\begin{eqnarray}
	\CS &=&      (\bar\psi_1                  \psi_2)
                       	(\bar\psi_3                  \psi_4)\,, 
\nonumber \\
	\CV &=&\sum_\mu (\bar\psi_1\gamma_\mu        \psi_2)
                       	(\bar\psi_3\gamma_\mu        \psi_4)\,. 
\nonumber \\
	\CT &=&\sum_{\mu<\nu}(\bar\psi_1\gamma_\mu\gamma_\nu  \psi_2)
                        (\bar\psi_3\gamma_\mu\gamma_\nu  \psi_4)\,, 
\label{eq:Bklstruct} \\
	\CA &=&\sum_\mu (\bar\psi_1\gamma_\mu\gamma_5\psi_2)
        	     	(\bar\psi_3\gamma_\mu\gamma_5\psi_4)\,, 
\nonumber \\
	\CP &=&      (\bar\psi_1\gamma_5          \psi_2)
                       	(\bar\psi_3\gamma_5          \psi_4)\,.
\nonumber 
\end{eqnarray}
Keeping the four flavors distinct eliminates ``penguin'' diagrams,
and allows only a single Wick contraction for each operator.
In the text, we use subscripts to indicate incomplete summation over
repeated Lorentz indices. For example, $\CA_s$ denotes that the sum over
$\mu$ is only over $1-3$ and not $4$, while $\CT_t$ indicates that $\nu=4$
in the tensor operator.
Each operator also comes with two possible sets of color indices,
which we distinguish with superscripts, e.g.
\begin{equation}
\CS^1 = (\bar\psi_1^a \psi_2^b) (\bar\psi_3^b \psi_4^a)\,\ \ {\rm and}\ \ 
\CS^2 = (\bar\psi_1^a \psi_2^a) (\bar\psi_3^b \psi_4^b)\,.
\end{equation}
The superscript is the number of loops that the color indices form
if we take a matrix element of the operators
between states created by $(\bar\psi_2 \Gamma\psi_1)$ and
$(\bar\psi_4\Gamma\psi_3)$.

In the following we use two bases for the Lorentz structures:
the ``original'' basis
\begin{equation}\
\CO'_i = \left[\CS, \CV, \CT, \CA, \CP\right] \,,
\end{equation}
and the ``practical'' basis
(related by a non-orthogonal transformation)
\begin{equation}
\CO_i = \left[(\CV+\CA), (\CV-\CA), -2(\CS-\CP), (\CS+\CP), -(\CS+\CP-\CT)/2
\right] \,.
\end{equation}
The advantages of each will become clear in the following.
In each basis there is a matrix which implements Fierz transformations.
This is simplest in the practical basis
\begin{equation}
\CF^{\rm\, prac} = \left( \begin{array}{rrrrr} 
	1 & 0 & 0 & 0 & 0 \\
	0 & 0 & 1 & 0 & 0 \\
	0 & 1 & 0 & 0 & 0 \\
	0 & 0 & 0 & 0 & 1 \\
	0 & 0 & 0 & 1 & 0 
	         \end{array} \right)\,.
\end{equation}
In either basis this satisfies $\CF^2=1$.
There is also a matrix which implements ``charge conjugation'',
which here means the effect of transposing the Dirac
matrices in one of the bilinears, and conjugating the result with
$C=\gamma_0\gamma_2$. This is simplest in the original basis,
\begin{equation}
\CC^{\rm orig} = {\rm diag}(\ 1, -1, -1,\ 1,\ 1) \,.
\end{equation}
Again, in either basis one has $\CC^2=1$.
Note that this definition is specific to
four dimensional Dirac matrices.

To combine the Lorentz and color indices we use the notation
%$\vec{\CO_i}$ (or $\vec{\CO'_i}$), 
%where $i$ runs over the five Lorentz structures, 
%and the vector corresponds to the two color structures
\begin{equation}
\stackrel{\longrightarrow}{\CO}_i = 
\left( \begin{array}{c} \CO_{i}^1\\ \CO_{i}^2 \end{array} \right) \,.
\end{equation}
Then, as shown in Ref.~\cite{SharpePatel}, one-loop matching has the form
\begin{equation}
\stackrel{\longrightarrow}{\CO}^{\rm cont}_i = 
\stackrel{\longrightarrow}{\CO}^{\rm lat}_i
+ {\lambda} \sum_j \left( 
  {\CM}^a_{ij} \stackrel{\longleftrightarrow}{C_a} 
+ {\CM}^b_{ij} \stackrel{\longleftrightarrow}{C_b} 
+ {\CM}^c_{ij} \stackrel{\longleftrightarrow}{C_c}  \right) 
\stackrel{\longrightarrow}{\CO}^{\rm lat}_j \,.
\label{eq:mastercorr}
\end{equation}
There are three types of diagram which contribute, denoted
$a$, $b$ and $c$, each giving rise to a color matrix $C$ and
a matrix acting on the Lorentz indices.
The color matrices are
\begin{equation}
 \stackrel{\longleftrightarrow}{C_a} 
	  = \frac16 \left( \begin{array}{rr} 
                   -1 & 3 \\ 0 & 8 \end{array} 
            \right)
\,,\quad
 \stackrel{\longleftrightarrow}{C_b} 
	  = \frac16 \left( \begin{array}{rr} 
                   8 & 0 \\ 3 & -1 \end{array} 
            \right)
\,,\quad
 \stackrel{\longleftrightarrow}{C_c}
          = \frac16 \left( \begin{array}{rr} 
                   -1 & 3 \\ 3 & -1 \end{array} 
            \right)
\,.
\end{equation}
The core of the calculation is of the $5\times5$ correction matrices $\CM$.

The first stage of this calculation is to determine $\CM$ in 
a scheme in which the gamma-matrices are four-dimensional, say DRED.
The ``a''-diagrams are just those in which the two bilinears in the
four-fermion operator are matched independently 
(see Ref.~\cite{SharpePatel}).
Thus $\CM_a$ is simple in the original basis:
\begin{equation}
\CM_a = 2\ {\rm diag}[c_S, c_V, c_T, c_A, c_P] \,.
\label{eq:fierzaa} \\
\end{equation}
%where all of the bilinear coefficients are in the DRED scheme.
Following Ref.~\cite{pertmarti}
one can then obtain $\CM_b$ and $\CM_c$ using Fierz 
and charge conjugation transformations:
\begin{eqnarray}
\CM_b &=& \CF \CM_a \CF \,, 
\label{eq:fierzab} \\
\CM_c &=& - \CC \CF \CM_a \CF \CC \,.
\label{eq:fierzca}
\end{eqnarray}
These results can then be inserted in the general formula
Eq.~(\ref{eq:mastercorr}).
Fierz symmetry also requires that $\CM_c$ be self-conjugate:
\begin{equation}
\CM_c = \CF \CM_c \CF \,.
\label{eq:fierzcc}
\end{equation}
This result, together with Eq. (\ref{eq:fierzca}), 
and the peculiar fact that $\CF \CC \CF \CC \CF = \CC$,
requires that $[\CC, \CM_a] = 0$.
This is trivially true given
that $\CM_a$ and $\CC$ are diagonal in the original basis.

This construction fails for NDR, however, because one is then
dealing with n-dimensional gamma matrices.
It is useful, nevertheless, to introduce an 
intermediate scheme, ${\rm NDR}'$, which is {\sl defined} by
Eq.~(\ref{eq:fierzaa}),
with NDR bilinear matching coefficients,
together with Eqs. (\ref{eq:fierzab}) and (\ref{eq:fierzca}).
That is, we enforce Fierz and charge-conjugation symmetries by hand.
%We stress that this is an intermediate scheme, which we do not know
%how to extend beyond one-loop. It is useful because the corrections
%can be written in terms of bilinear matching coefficients, and because
%it is close to NDR.

The second stage is the extension of the calculation to NDR.
For this, we need to define the continuation of the four-fermion operators to
$n$-dimensions. There are many ways to do this (corresponding to different
choices of ``evanescent'' operators), and we wish to make a choice which
maintains the Fierz relations Eqs. (\ref{eq:fierzab}) and (\ref{eq:fierzcc}).
We have not found a way of simultaneously maintaining Eq. (\ref{eq:fierzca}),
and suspect that this is not possible. 
Thus, in NDR, $\CM_b$ can be obtained from $\CM_a$, 
while $\CM_c$ is independent.
Fierz-symmetric choices for $(\CV\pm\CA)$ and $(\CS-\CP)$ were given in
Refs. \cite{Burasetal,Ciuchinietal}, and we use these.
Definitions for $(\CS+\CP)$ and $\CT$ were given in Ref. \cite{SharpePatel}, 
but the choices made there did not maintain Fierz symmetry.
We have found Fierz-symmetric choice, which we now explain.

We work in the ``practical'' basis.
The operators are extended to $n=4-2\epsilon$ dimensions such
that they are the even parity parts of
\begin{eqnarray}
\CO_1 &=& (\bar\psi_1 \gamma_\mu L \psi_2)\,
	  (\bar\psi_3 \gamma_\mu L \psi_4) \,,\\
\CO_2 &=& (\bar\psi_1 \gamma_\mu L \psi_2)\,
	  (\bar\psi_3 \gamma_\mu R \psi_4) \,,\\
\CO_3 &=& -2 (\bar\psi_1 L \psi_2)\,
	  (\bar\psi_3 R \psi_4) \,,\\
\CO_4 &=& 	(\bar\psi_1 L \psi_2)\,
		(\bar\psi_3 L \psi_4) \,,\\
\CO_5 &=& -\nfrac18 (1 + 3 \epsilon/4)\ 
(\bar\psi_1 \gamma_\mu \gamma_\nu L \psi_2)\,
(\bar\psi_3 \gamma_\nu \gamma_\mu L \psi_4)
\,.
\end{eqnarray}
Here we are implicitly summing repeated indices over all $n$ values,
and defining $L=(1-\gamma_5)$ and $R=(1+\gamma_5)$.
The odd parity parts of these operators are not important for the
following discussion, but we include them for generality.
The analysis is unchanged if we change the sign of $\gamma_5$.
The peculiar factor of $(1+3\epsilon/4)$ in $\CO_5$ is necessary
in order that, in the final result, $\CM_c$ is Fierz self-conjugate.
When we calculate one-loop corrections in NDR the resulting operators
have the form
\begin{equation}
\CO = \sum_{i=1,5} d_i \CO_i + {\rm evanescent\ operators}\,,
\end{equation}
where the evanescent operators vanish when $n=4$.
We determine the $d_i$ using projection operators.
Fierz-symmetric projectors onto $\CO_{1-3}$ have been given in
Refs. \cite{Burasetal,Ciuchinietal}.
To determine $d_4$ and $d_5$ [up to terms of $O(\epsilon^2)$] we use
\begin{equation}
\left(\begin{array}{r} d_4 \\ d_5 \end{array}\right) =
-\frac{1+5\epsilon/12}{3}
\left(\begin{array}{cc} 
 (1+\epsilon/4)		& 2(1-\epsilon/4) \\
 2			& 1
\end{array}\right) 
\left(\begin{array}{r} P_4(\CO) \\ P_5(\CO) \end{array}\right) \,.
\label{eq:d45}
\end{equation}
where, in the notation of Ref. \cite{Ciuchinietal}, the projectors are
\begin{eqnarray}
P_4 &=& \frac1{32}L \otimes L \ \ {\rm and} \\
P_5 &=& -\frac1{256} (\frac4{n})^2
%\sum_{\mu,\nu=1}{n} 
\gamma_\mu \gamma_\nu L \otimes \gamma_\nu \gamma_\mu L \,.
\end{eqnarray}
The matrix in Eq. (\ref{eq:d45}) has been determined by requiring
$d_i=1$ if $\CO=\CO_i$, $i.e.$ the basis operators project back onto themselves.

Using these definitions, we have calculated the matching between
NDR and ${\rm NDR}'$, and thus between NDR and lattice operators.
We find for our final results
\begin{eqnarray*}
\CM_a&=& \left( \begin{array}{ccccc} 
 c_{V+A} 	& c_{V-A} 	& 0 		& 0		& 0 \\
 c_{V-A}	& c_{V+A}	& 0		& 0		& 0 \\
 0      	& 0		& c_{S+P}       & -2 c_{S-P}	& 0 \\
 0		& 0		&-\frac12c_{S-P}& c_{S+P}  	& -1 \\
 0		& 0 		& \frac14c_{S-P}&\frac12(2c_T-c_{S+P}) 
								& 2 c_T
	         \end{array}\right) \,, 
\\[2em]
%\CM_b &=& \CF \CM_a \CF \,, 
%\\[2em]
\CM_c &=& \left(  
\begin{array}{ccccc} 
-(c_{S+P}+6)	& 0		& 0		& 2 c_{S-P}  & 2c_{S-P} \\
  0		& -(c_{V+A}+6)  & c_{V-A}	& 0	     & 0	\\
  0		& c_{V-A}	& -(c_{V+A}+6)  & 0	     & 0	\\
\frac14 c_{S-P}	& 0		& 0	& -\frac12(c_{S+P}+2c_T+4)
						& -\half(c_{S+P}-2c_T+2) \\
\frac14 c_{S-P}	& 0	 	& 0	& -\frac12(c_{S+P}-2c_T+2)
						&-\frac12(c_{S+P}+2c_T+4)
	         \end{array}\right) \,,
\end{eqnarray*}
where all the $c_i$ are in the NDR scheme, and we have defined
$c_{S+P}=c_S+c_P$, etc.
$\CM_b$ continues to be determined from Eq.~(\ref{eq:fierzab}).
Note that $\CM_c$ is Fierz self-conjugate. 
The results for ${\rm NDR}'$ and DRED can be obtained by deleting the
numerical constants, i.e. keeping only the terms involving the $c_i$. 

\bigskip

We conclude by presenting the explicit results needed for 
the matrix elements we study in this paper.
The operators of interest are
\begin{eqnarray}
\CQ &=&	\half \big[ (\bar s_a \gamma_\mu L d_a) (\bar s_b \gamma_\mu L d_b) +
		    (\bar s_a \gamma_\mu L d_b) (\bar s_b \gamma_\mu L d_a) 
\big] \,, 
\label{eq:qdef}\\[0.2em]
\CQ_7^{3/2} &=& (\bar s_a \gamma_\mu L d_a) \  \big[ 
	 	(\bar u_b \gamma_\mu R u_b) - (\bar d_b \gamma_\mu R d_b) \big]
	     +  (\bar s_a \gamma_\mu L u_a) (\bar u_b \gamma_\mu R d_b) \,, 
\label{eq:q7def}\\
\CQ_8^{3/2} &=& (\bar s_a \gamma_\mu L d_b) \  \big[ 
	 	(\bar u_b \gamma_\mu R u_a) - (\bar d_b \gamma_\mu R d_a) \big]
	     +  (\bar s_a \gamma_\mu L u_b) (\bar u_b \gamma_\mu R d_a) 
\,,\label{eq:q8def}\\[0.2em]
\CQ_{S} &=& \half \big[	(\bar b_a L s_a) (\bar b_b L s_b) 
- \nfrac18 (1 + 3 \epsilon/4) 
(\bar b_a \gamma_\mu\gamma_\nu L s_b) (\bar b_b \gamma_\nu\gamma_\mu L s_a)
\big] \,. 
\label{eq:qsdef}
\end{eqnarray}
Following Ref.~\cite{JaminPich},
we have extended $\CQ$ and $\CQ_S$ to $n$-dimensions in such a way
that the Fierz symmetry is explicit.
None of these operators have penguin contractions (in the isospin
symmetric limit), and so all can be matched onto the lattice using the
approach explained above.
%We stress that we are not including the effects of finite quark masses.

The first step is to rewrite the operators using the four flavor notation.
For this purpose it is useful to introduce a third operator basis,
that of Fierz self-conjugate and anti-conjugate operators:
\begin{equation}
\CO^\pm_i = [\
	(\CO_1^2\pm\CO_1^1),\
	(\CO_2^2\pm\CO_3^1),\
	(\CO_3^2\pm\CO_2^1),\
	(\CO_4^2\pm\CO_5^1),\
	(\CO_5^2\pm\CO_4^1)] \,.
\end{equation}
The operators of interest can be written in this new basis as
\begin{eqnarray}
\CQ &\longrightarrow&  2 \CO^+_{1}\,,\\
\CQ_7^{3/2} &\longrightarrow&  \CO^+_{2}\,,\\
\CQ_8^{3/2} &\longrightarrow&  \CO^+_{3}\,.\\
\CQ_S &\longrightarrow& 2 \CO^+_{4}\,,
\end{eqnarray}
where the factors of two come from the additional Wick contractions for 
$\CQ$ and $\CQ_S$ compared to the four flavor operators.
Fierz symmetry forbids mixing between the $\CO^+_{i}$ and $\CO^-_{i}$,
and so the matching coefficient falls into two $5\times5$ blocks.
We give the results for the Fierz self-conjugate operators:
\begin{equation}
\CO^+_i({\rm NDR}) - \CO^+_i({\rm lat}) =
\lambda\, c^+_{ij}\, \CO^+_j({\rm lat})
\label{eq:defncij}
\end{equation}
with
\begin{equation}
c^+_{ij} =
\left(\scriptstyle{\begin{array}{rrrrr} 
 Z_+ 	& 0 	& 0 	& 0 	& 0 \\
 0 	& Z_1	& Z_{78}& 0	& 0 \\
 0	& Z_{87}& Z_2	& 0	& 0 \\
 0	& 0	& 0	& Z_4	& Z_{45} \\
 0	& 0	& 0	& Z_{54}& Z_5
       \end{array}} \right) 
+ {Z^* \over 24} 
\left( \scriptstyle{ \begin{array}{rrrrr} 
 0 	& -22 	&  2 	& 16 	& 16 	\\
-16	&  -6	&  2	&-24	&  8	\\
 -4	&   2	& -6 	&-64	&  0 	\\
  2     &  -1	&-13	&  0	&  0	\\
  2	&   2	&  2	&  0	&  0
       \end{array}} \right) \,.
\label{eq:defnZ}
\end{equation}
The expressions and numerical results
for the $Z$'s are collected in Table \ref{tab:ffcorr}.
Our results in the DRED scheme for the matching coefficients
of $\CO_{1+}$, $\CO_{2+}$ and $\CO_{3+}$ agree with those of
Ref.~\cite{pertmarti}. 
The results of Ref. \cite{Bk93LANL} for $Z_+$, $Z_1$ and $Z_2$ 
using NDR are, however, wrong, and are corrected by those given here.

\begin{table}
\renewcommand{\arraystretch}{1.2}
\begin{center}
\begin{tabular}{lclc}
\hline
	& NDR result	& $\ \ $NDR value & DRED $-$ NDR  \\
\hline
$Z_+$	& $\nfrac13 (5 c_V + 5 c_A - c_S - c_P) - 2$ 
		& $-50.841 - 4 \ell + 2\, {\rm tad}$	& $3$ 
\\
$Z_1$	& $\nfrac16 (9 c_V + 9 c_A - c_S - c_P) + 1$
		& $-47.698 - 2 \ell + 2\, {\rm tad}$	& $\nfrac16$
\\
$Z_2$	& $\nfrac43 (c_S + c_P) + 1$
			& $-46.398 + 16 \ell + 2\, {\rm tad}$& $\nfrac53$
\\
$Z_{78}$& $\nfrac12 (c_S + c_P - c_V - c_A) - 3$
			& $\ -2.567 + 6 \ell$		& $\nfrac72$
\\
$Z_{87}$& $-3	$	& $\ -3	$			& $3$
\\
$Z_4$  	& $\nfrac1{12} (10 c_T + 11 c_S + 11 c_P - 2)$
		& $-46.894 + \nfrac{28}3 \ell +2\,{\rm tad}$&$\nfrac76$
\\
$Z_5$	& $\nfrac13 (10 c_T - c_S -c_P - 2)$
			& $-45.384 -\nfrac{32}3 \ell + 2\,{\rm tad}$	
							&$-\nfrac{10}3$
\\
$Z_{45}$& $\nfrac1{12} (2 c_T - c_S - c_P - 26)$
			& $\ -2.033 - \nfrac43 \ell 	$	& $2$
\\
$Z_{54}$& $\third (2 c_T - c_S - c_P - 2 )$
			& $\ -0.131 - \nfrac{16}3 \ell$ 	& $-\nfrac43$
\\
$Z^*$ 	& $c_S-c_P = 2 (c_A-c_V)$ 
			& $\ \ 9.644 			$& $0$
\\
\hline
\end{tabular}
\caption{Matching coefficients, as defined in Eqs.~\ref{eq:defncij} and \ref{eq:defnZ},
needed for Fierz self-conjugate four-fermion operators.
The expressions are given for the NDR scheme and are valid for
any local lattice operator. Expressions for
DRED are obtained by keeping only the $c_i$ terms.
The numerical values are for the NDR scheme with unimproved Wilson
lattice fermions, with $\ell = \log(\mu a)$, 
and ${\rm tad}=17.14$ being the effect of tadpole improvement.
The last column gives the difference between the
numerical values for DRED and NDR schemes.}
\label{tab:ffcorr}
\end{center}
\end{table}

The matrix elements needed in the study of B-lifetimes are those of 
operators $\CO_4^+$ and $\CO_{5}^+$.
Since these are new, we write them in a form which is more directly
practical.
In the 4-flavor notation these operators are 
\begin{eqnarray}
\CO_{4}^+ &=& (\CS^2+\CP^2) - \half (\CS^1+\CP^1 - \CT^1) \,, \nonumber \\
\CO_{5}^+ &=& (\CS^1+\CP^1) - \half (\CS^2+\CP^2 - \CT^2) \,.
\end{eqnarray}
The 1-loop continuum operators, in the NDR scheme, are 
\begin{eqnarray}
\CO_{4}^+({\rm cont}) &=& 
       [1 + \lambda (-46.894 + 2 {\rm tad} + 28/3 \ln(\mu a))]  
\CO_{4}^+({\rm lat}) \nonumber \\
&{}& + \lambda (-2.033 - 4/3 \ln(\mu a))
        \CO_{5}^+({\rm lat}) \nonumber \\
&{}& + {\lambda Z^* \over 24} [2 \CS^1 + 26 \CS^2 -2\CP^1 -26\CP^2 
             -11 \CV^1 + \CV^2 + 15 \CA^1 + 3 \CA^2] \,, \nonumber \\
\CO_{5}^+({\rm cont}) &=& 
       [1 + \lambda (-45.384 + 2 {\rm tad} - 32/3 \ln(\mu a))]
  \CO_{5}^+({\rm lat}) \nonumber \\
&{}& + \lambda (-0.131 - 16/3 \ln(\mu a))
  \CO_{4}^+({\rm lat}) \nonumber \\
&{}& + {\lambda Z^* \over 6} 
[ (\CP^1 - \CS^1) + (\CP^2 - \CS^2) + \CV^1 + \CV^2] \,.
\label{e:O45defs}
\end{eqnarray}
The $B$ parameters are obtained by dividing the above by their VSA forms 
\begin{eqnarray}
\CO_{4}^+({\rm VSA}) 
	&=& (+5/6) [1+ \lambda Z_P]^2 (P \otimes P) \,, \nonumber \\
\CO_{5}^+({\rm VSA}) 
	&=& (-1/6) [1+ \lambda Z_P]^2 (P \otimes P) \,.
\label{e:VSA45defs}
\end{eqnarray}
These two operators mix under the renormalization group flow. The
1-loop anomalous dimension matrix governing this flow is
\[
\left(
\begin{array}{cc}
-28/3   &  4/3  \\
 16/3   &  32/3 
\end{array} 
\right)
\]
and the 2-loop contribution is yet to be calculated.  Consequently,
we state our results for B-parameters after the 1-loop matching at scale $\mu=q^*$, 
$i.e.$ without any running in the continuum which requires the 2-loop results.

%%%%%%%%%%%%%%%%%%%%%%%%%%%%%%%%%%%%%%%%%%%%%%%%%%%%%%%%%%%%%%%%%%

\end{document}